\documentclass[11pt,journal,onecolumn]{IEEEtran}
\usepackage{amssymb}
\usepackage{amstext}
\usepackage{latexsym}
\usepackage{times}
\usepackage{amssymb}
\usepackage{amsmath}
\usepackage{epsfig}
\usepackage{cite}
\usepackage{verbatim}
\usepackage{enumerate}
\usepackage{subfigure}
\usepackage{multicol}
\usepackage{pstricks}
\usepackage{pst-node}
\usepackage{setspace}
\usepackage{graphicx}
\usepackage{picinpar}

\newtheorem{theorem}{\bf Theorem}

\begin{document}
\title{The Effect of Noise Correlation in AF\\ Relay Networks\\}
\author{\authorblockN{Krishna S. Gomadam, Syed A. Jafar}\\
\authorblockA{Electrical Engineering and Computer Science\\
University of California, Irvine, CA 92697-2625\\ Email: \{kgomadam,
syed\}@uci.edu}\\} \maketitle \IEEEpeerreviewmaketitle

\begin{abstract}
In wireless relay networks, noise at the relays can be correlated
possibly due to common interference or noise propagation from
preceding hops. In this work we consider a parallel relay network
with noise correlation. For the relay strategy of
amplify-and-forward (AF), we determine the optimal rate maximizing
relay gains when correlation knowledge is available at the relays.
The effect of correlation on the performance of the relay networks
is analyzed for the cases where full knowledge of correlation is
available at the relays and when there is no knowledge about the
correlation structure. Interestingly we find that, on the average,
noise correlation is beneficial regardless of whether the
relays know the noise covariance matrix or not. However, the
knowledge of correlation can greatly improve the performance.
Typically, the performance improvement from correlation knowledge
increases with the relay power and the number of relays. With
perfect correlation knowledge the system is capable of canceling
interference if the number of interferers is less than the number of
relays.

For a dual-hop multiple access parallel network, we obtain closed
form expressions for the maximum sum-rate and the optimal relay
strategy. The relay optimization for networks with three hops is
also considered. For any relay gains for the first stage relays,
this represents a parallel relay network with correlated noise.
Based on the result of two hop networks with noise correlation, we
propose an algorithm for solving the relay optimization problem for
three-hop networks.
\end{abstract}
\vfill
\newpage
\section{Introduction}
Wireless mesh networks where information is transferred via multiple
hops and routes provide significant throughput enhancement and have
been the focus of much recent work \cite{viswanthan_mukherjee,
mac_bc_dual, relay_based_deployment, amplify_borade}. Various relay
strategies have been studied in literature \cite{cap_theorems_cover,
cap_theorems_relay_kramer, laneman_efficientprotocols}. These
strategies include amplify-and-forward \cite{laneman_worell_exploit, laneman_mod_det}, where the relay sends a scaled version of
its received signal to the destination, demodulate-and-forward
\cite{laneman_mod_det} in which the relay demodulates individual
symbols and retransmits, decode-and-forward
\cite{laneman_efficientprotocols, cap_theorems_cover} in which the
relay decodes the entire message, re-encodes it and re-transmits it
to the destination, and compress-and-forward
\cite{cap_theorems_cover, cap_theorems_relay_kramer} where the
relay sends a quantized version of its received signal. Among these
strategies, amplify and forward has been found to be highly suitable
for parallel relay networks for its ability to pass on soft
information \cite{Gomadam2006, Gomadam2006a, jsac_memoryless}. In
one of the first works \cite{schein_gallager} on the parallel relay
channel, it is shown that AF achieves the capacity cut-set bound at
high relay power. In \cite{amplify_borade} it is shown that full
degrees of freedom can be achieved with AF in a multi-hop parallel relay
network where the hops are orthogonalized by design. There are
numerous other examples in literature that corroborate the
effectiveness of AF \cite{cap_lar_gauss_relay, elgamaal_linear,
relay_without_delay, jsac_memoryless, Bolcskei2006a, chen-2007,
break_net}. For an AF relay network, the relay design involves
optimizing the relay amplification factors to maximize performance.
Previous work pertaining to relay optimization for AF relay networks
includes both single-user relaying \cite{larsson_hong, maric_yates,
distributed_mmse_nima_sayed, parallel_yener, Yi2007, zhao_adve} and
multi-user relaying \cite{mac_bc_dual,
distributed_multiuser_mmse_berger_wittenben}.

Most work in the literature \cite{larsson_hong,nabar_jsac03,
maric_yates, distributed_mmse_nima_sayed, parallel_yener,
distributed_multiuser_mmse_berger_wittenben, Yi2007,zhao_adve} assumes
independent noise at the relay terminals. However, noise correlation
between nodes occurs in wireless relay networks due to several
reasons. In this paper we explore the effect of noise correlation for the following two models.
\begin{figure}[h]
\center
\input{common_int.pstex_t}
\caption{\textsf{Noise correlation model 1:~~Parallel relay network
with common interference }}\label{fig:common_int}
\end{figure}
\begin{enumerate}
\item{{Common Interference Model}}\\
Due to the broadcast nature of wireless networks, the relays are
exposed to a set of common interferers resulting in correlated noise
at the nodes. As shown in Fig. \ref{fig:common_int}, each relay in
addition to its local noise observes common interference.

\item{{Noise propagation with multi-hop AF relaying (Three-hop model)}}\\
Consider a three-hop network as shown in Fig. \ref{fig:multi_hop}.
If the first stage relays amplify their received signal, each relay
in the second stage observes a linear combination of noise from the
preceding stage relays along with its own local noise. This clearly
results in correlated noise at the second-stage relays.
\begin{figure}[h]
\center
\input{multi_hop.pstex_t}
\caption{\textsf{Noise correlation model 2:~~Three-hop parallel
relay network}}\label{fig:multi_hop}
\end{figure}
\end{enumerate}
Both the above models are of considerable practical importance. The
natural question is whether the relays can exploit the correlation
structure to improve performance. In practice, learning correlation
may result in network overheads. Whether such overheads are
justified depends on the potential advantage of learning
correlation. Thus, our goal in this work is to estimate the
improvement in performance when perfect correlation knowledge is
available at the relays. The optimal relay design in this case will
have the following two objectives:
\begin{enumerate}
\item{Increase signal power:} The relay gains can be designed such that the copies of desired signal adds up in phase.
\item{Reduce interference power: } The relay gains may be chosen such that the common interference terms add out of phase.
\end{enumerate}
In general, these objectives cannot be achieved simultaneously and
the relay design can be expected to be a trade-off between the two
objectives.

This paper also addresses the relay optimization problem for
networks with three hops. Such a network is interesting as it
combines multiple hops with multiple paths. The problem of routing
in the traditional link based decode and forward involves selecting
a relay at each hop that maximizes the end to end throughput between
the original source and the final destination. If there are $N$
relays in each of the $M$ hops, the problem is to find the best
route among $N^{(M-1)}$ routes. With AF, all the routes are utilized
with priority for each route determined by the power allocation and phase equalization at
each of the relays across multiple hops. As we can control
the priority for the routes, optimizing relay gains may be viewed as \emph{soft-routing}.

\subsection{Results}
The main results of this paper are summarized below.
\begin{enumerate}
\item We obtain closed form solutions for the optimal relay
amplification vector and the maximum sum rate for the network with
correlated noise at the relays. This result generalizes the
single-user relay optimization in \cite{maric_yates, larsson_hong}
and the multi-source relay optimization in \cite{mac_bc_dual} both of which assume independent and identically distributed (i.i.d.) noise at the relays.
\item{We find that correlation, on the average, is always beneficial regardless of the presence or absence of  correlation knowledge at the relays.
This is true irrespective of the channel state information (CSI) at the relays.}
\item We study the benefits of exploiting
noise correlation at the relays. We compare the maximum sum rate
without correlation knowledge (relays use optimal amplification
factors based on uncorrelated noise assumption even if noise is
correlated) versus capacity with correlation knowledge.  The
following key questions are answered: Does correlation help? Is the
correlation knowledge more (or less) helpful as the number of relays
increases? What is the effect of correlation as the power at the
relays increases, and when the first hop/second hop becomes
stronger? We also provide asymptotic results to characterize the impact of relay
noise correlation.
\item We apply the results of two hop relay networks with correlated
noise to solve the relay optimization problem for three-hop AF relay
networks with independent noise processes at the relays. We also
characterize the behavior of the three-hop network as the transmit
power at any of the stages tends to $\infty$. When the first or the
last stage power is very high, the network can be reduced to a
two-hop channel. When both the first and last stage powers are high,
the network can be reduced to a point to point MIMO channel with a
single data stream.
\end{enumerate}
\subsection{Notation}
We use bold upper letters to denote matrices and bold lower letters
to denote vectors. Further $(.)^{*}$, $(.)^T$, $(.)^{\dagger}$ stand
for conjugation, transposition and Hermitian transposition,
respectively. ${\bf a} \odot {\bf b}$ represents element wise
multiplication of two vectors. ${\bf A}=\mbox{diag}({\bf a})$
denotes a diagonal matrix {\bf A} that contains on its diagonals the
elements of ${\bf a}$.  $\mbox{det}(\bf A)$, $\mbox{Tr}(\bf A)$,
$\lambda_{\max}({\bf A})$, and ${\bf v}_{\max}({\bf A})$ denote the
determinant, trace, principal eigen-value, and principal
eigen-vector of ${\bf A}$ respectively. $\bf A^{-\dagger}$ denotes the Hermitian transposition of $\bf A^{-1}$. ${\bf v}(i)$ denotes the
$i^{th}$ element of the vector $\bf v$ while ${\bf A}_{ij}$ denotes the element in the $i^{th}$ row and $j^{th}$ column of the matrix $\bf A$. $\mathbb{E}$ stands for the expectation
operator. ${\bf v} \sim \mathcal{CN}({\bf w},{\bf A})$ indicates $\bf v$ is a complex Gaussian column vector with $\mathbb{E}[{\bf v}]={\bf w}$ and $\mathbb{E}[{\bf vv^{\dagger}}]={\bf A}$.

This paper is organized as follows. In the next section we consider
a two hop relay network with noise correlation and address the relay optimization problems for both single and multi-user scenarios. We analyze the impact of correlation in Section \ref{Section:help_hurt}.
In Section \ref{Section:3hops} we address the relay optimization problem for three-hop AF networks. We conclude with Section \ref{Section:conclusion}.

\section{Two-hop parallel relay network with noise correlation}
\begin{figure}
\center
\input{single_corr.pstex_t}
\caption{\textsf{Two-hop single user parallel relay
network}}\label{single_corr}
\end{figure}

\subsection{System Model}
We consider a two-hop parallel relay network as shown in Fig.
\ref{single_corr}. In this model, the source $S$ communicates to the
destination $D$ through a set of $N$ half-duplex relays, $R_{i}$,
$i=1...N$. The data transmission takes place in two time slots. In
the first slot, the source transmits to the relays and in the second
slot, all the relays simultaneously forward their received signal to the destination. Note that there is no direct link between the source and the
destination. All the nodes are equipped with a single antenna. The $N \times 1$ channel between the source and the relays is
$\bf f$ while $\bf g$ is the $N \times 1$ channel between the relays and the
destination. The entries of $\bf f$ and $\bf g$ are independent and i.i.d. complex Gaussian random variables with zero mean and unit variance. The source has power $P$ and the
relays have a total power of $P_{R}$.  The relay received symbols during the
first time slot are
\begin{equation}
{\bf r}={\bf f}x+{\bf n_{R}}
\end{equation} where $x$ is the source transmitted signal with power $\mathcal{E}[|x|^2]=P$, and ${\bf n_{R}} \sim \mathcal{CN}({\bf 0,~ K})$ is AWGN with
the covariance matrix $\bf K$, given by \[{\bf K}=\mathbb{E}[{\bf
n_{R}n_{R}^{\dagger}}].\]
In the second slot, the relay $R_i\}_{i=1}^{N}$ scales its received
signal with a complex scaling factor $d_i$ and transmits to the
destination. We collect the relay gain factors $d_i$ in an $N \times
1$ vector $\bf d$.  Then the received signal at the destination can
be expressed as
\begin{equation}
y={\bf d^{T}Gf}x+{\bf d^{T}Gn_{R}}+n_{D} \label{received_signal}
\end{equation} where
${\bf G}=\mbox{diag}({\bf g})$ and $n_{D} \sim \mathcal{CN}(0, 1)$
is additive white Gaussian noise (AWGN). Further, we assume that there is no correlation between
destination noise $n_{D}$ and the relay noise vector $\bf n_{R}$ as
these noise processes occur at two different time slots. We denote this relay
network with the shorthand notation $\mathfrak{R}_{2}(P, {\bf f},
{\bf d}, P_{R}, {\bf g})$ where the subscript 2 indicates that the
network consists of two hops. The sum power constraint of the relays can be expressed as
\begin{equation}
{\bf d^{\dagger}}\left[\left({\bf
ff^{\dagger}}P+{\bf K}\right)\odot {\bf I}\right]{\bf d}=P_{R}
\label{power_constraint}
\end{equation} where $\odot$ represents the
Hadamard product or the element-wise multiplication. Note that there
is a strict equality for the power constraint. This is because higher total relay transmit power is equivalent to lower noise power at the destination. Finally a
word about the sum-power constraint. With individual
power constraints, not all the relays may operate at full power
\cite{jing_jafarkhani}. Thus the total power expenditure at the
relays may change depending on the channel conditions. With
sum-power constraint relays use all the power.
This is helpful in interference limited systems where the
interference from the relays to other wireless links in the network
needs to be estimated. And most importantly, the sum power
constraint allows tractable analysis and useful insights can be
obtained as a result of it.
\subsection{Relay Optimization} From the destination received signal in (\ref{received_signal}), the
signal-to-noise-ratio (SNR) at the destination is given by
\begin{equation}\mbox{SNR}=\frac{\left|{\bf d^{T}G}{\bf f}\right|^2P}{{\bf
d^{T}GKG^{\dagger}d^{*}}+1}\end{equation}  and the corresponding transmission
rate\footnote{Gaussian inputs at the source are optimal for AF relays.} is
$R=\log(1+\mbox{SNR})$. We seek to maximize the transmission rate
with respect to the amplification vector ${\bf d}$ subject to the
sum power constraint at the relays. The following theorem provides
the optimal relay design and maximum
achievable rate with amplify and forward relays in a two hop parallel relay network with noise correlation.\\
\begin{theorem}
\label{THEOREM:SINGLE USER MAX RATE} The maximum achievable rate in
the two-hop parallel AF relay network $\mathfrak{R}_{2}(P, {\bf f},
{\bf d}, P_{R}, {\bf g})$ where the relay noise covariance matrix is $\bf K$, is given by
\begin{equation}
R^{\circ}=\log(1+\mbox{SNR}^{\circ})
\end{equation}
\begin{equation}
\mbox{SNR}^{\circ}=PP_{R}({\bf f\odot g})^{\dagger} {\bf A^{-1}}(
{\bf f\odot g}) \label{optimal_snr_dual_hop}
\end{equation} and
the optimal relay amplification vector ${\bf d}^{\circ}$ is
\begin{equation}
{\bf d}^{\circ}= \kappa \left({\bf A^{-1}(f \odot g)}\right)^{*}
\label{corr_amp_vector}
\end{equation}
where \begin{equation}{\bf A}= {\bf K\odot gg^{\dagger}}P_{R}+{\bf
ff^{\dagger}} P \odot {\bf I +K \odot I}\end{equation} and
\begin{equation}\kappa^2=\frac{P_R}{({\bf f\odot g})^{\dagger} {\bf
A^{-1}}\left[\left({\bf ff^{\dagger}}P+{\bf K}\right)\odot {\bf
I}\right]{\bf A^{-\dagger}}({\bf f\odot g})}.\end{equation}
\end{theorem}
The proof is presented in the Appendix.
\subsubsection*{Remark}In the absence of correlation, $\bf A$ is diagonal and
the phase of relay $R_i$ is such that it cancels the phase of the
forward channel ${\bf g}(i)$ and the backward channel ${\bf f}(i)$. However, with noise correlation, ${\bf A}$  is not a diagonal matrix. Therefore the optimal relay gain may not co-phase the input signal
as part of noise can be canceled with relay phase and gain
adjustments.

\subsection{Properties of the optimal relay network}
In this section, we consider the limiting properties of the optimal
relaying scheme. The proofs are available in the Appendix.
\label{SUBSECTION: PROPERTIES PROOF DUAL HOP}
\begin{enumerate}
\item{Relay power $P_{R}\rightarrow \infty$}:~
When the relay power is high, the network is equivalent to a single-input-multiple-output (SIMO)
system modeled by ${\bf y=f}x+{\bf n}$ where $\bf K$ is the noise
covariance matrix at the multiple antenna receiver and $P$ is the
transmit power of the single antenna transmitter. Its capacity is
given by
\begin{equation}
C=\log(1+{\bf f^{\dagger}K^{-1}f}P). \label{mrc_bound}
\end{equation}
Note that the system is independent of ${\bf g} $.
Here $\bf d^{\circ}=\kappa (G^{-\dagger}K^{-1}f)^{*}$ achieves (\ref{mrc_bound}).
When ${\bf K}$ is not invertible (or equivalently $\mbox{det}({\bf
K})=0$), the capacity is infinite. This means that the effective
noise at the destination can be completely eliminated regardless of
the source transmission rate. For any relay gain vector, $\bf d$,
the network is equivalent to a SIMO system
with the receive combining vector given by $({\bf d \odot g})$.\\
\item{Source power $P \rightarrow \infty$}:~
For the optimal ${\bf d}^{\circ}$, the network is equivalent to a MISO
system represented by $y={\bf gx}+n$ where $n$ is unit variance
AWGN. Its capacity is given by
\begin{equation}
C=\log(1+{\bf g^{\dagger}g}P_{R}), \label{mrt_bound}
\end{equation}
which is independent of the relay noise covariance matrix $\bf K$.
${\bf d}(i)=\kappa {\bf g^{*}}(i)/{\bf f}(i)$ achieves
(\ref{mrt_bound}) in the relay network. For a general $\bf d$, the
relay network represents a multiple-input-single-output (MISO)
system $y={\bf gx}+n$, with the beamforming vector $\bf \frac{(d
\odot f)}{\sqrt{(d \odot f)^{\dagger}(d \odot f)}}$ and transmit
power
equal to $P_{R}$. \\
\item{Singular $\bf K$ or $\mbox{det}({\bf K})=0$}:~
Singular\footnote{Theorem 1 holds for any noise covariance matrix
and we do not assume any structure for the covariance matrix.} $\bf
K$ in any of the noise correlation models of common interference and
three-hop network may be unrealistic as the relays have independent
local noise components. However the case of singular $\bf K$
provides useful insights when the power of the interference is very
high for the common interference model, and when the total power of
the first stage relays is very high in the three-hop model. Notice
that the rank of $\bf K$ indicates the number of independent noise
sources at the relays. Thus for singular $\bf K$ at most
($N-\mbox{rank}({\bf K})$) noise terms can be canceled with
appropriate relay gains. Due to destination noise, the capacity of
the relay network is finite even when $\bf K$ is singular.
\end{enumerate}

The relay optimization result of Theorem 1 can be extended to several multi-user scenarios. As an example we consider the multiple access relay channel in the following.
\subsection{Extension: Multiple access parallel relay network with noise correlation}
Consider a multi-source parallel relay network as shown in Fig.
\ref{multi_user_parallel_relay_network}. In this model, $L$ source
nodes wish to communicate to a common destination  with the
assistance of $N$ AF relays. In the first slot all the sources transmit.
The relay received signals are given by
\begin{equation}
{\bf r}=\sum_{i=1}^{L}{\bf f}_{i}x_{i}+{\bf n_{R}}
\end{equation}
where $x_i$ is the signal transmitted by the $i^{th}$ source whose
power is $\mathbb{E}[|x_i|^2]=P_{i}$. As in the single user case,
the relays scale their received signal and transmit to the
destination in the second slot. The received signal at the
destination can be expressed as
\begin{equation}
y=\sum_{i=1}^{L}{\bf d^{T}Gf}_{i}x_{i}+{\bf d^{T}Gn_{R}}+n_{D} \label{received_signal_mac}
\end{equation}
The following theorem provides the
maximum achievable sum rate and the optimal relay design.
\begin{figure}
\center
\input{corr.pstex_t}
\caption{\textsf{Two-hop multi-source parallel relay
network}}\label{multi_user_parallel_relay_network}
\end{figure}
\begin{theorem}
\label{THEOREM: MULTIUSER MAX RATE} The maximum achievable sum rate
in a two-hop multi-source parallel AF relay network with noise
correlation is given by
\begin{equation}
R^{\circ}=\log(1+\mbox{SNR}^{\circ})
\end{equation}
\begin{equation}
\mbox{SNR}^{\circ}=P_{R}\lambda_{\max}\left[{\bf A^{-1}B}\right]
\end{equation} and the optimal relay amplification vector $\bf d_{\Sigma}^{\circ}$ is
\begin{equation}
{\bf d_{\Sigma}^{\circ}}=\kappa \left({\bf v_{\max}\left[ A^{-1}B\right]}\right)^{*}
\end{equation}
where
\begin{equation}{\bf A}= {\bf K\odot gg^{\dagger}}P_{R}+\sum_{k=1}^{L}{\bf
f_{k}f_{k}^{\dagger}} P_{k} \odot {\bf I +K \odot I}\end{equation}
\begin{equation}{\bf B}=
\sum_{k=1}^{L}{\bf( f_{k}\odot g)( f_{k}\odot
g)^{\dagger}}P_{k}
\end{equation}
and $\kappa$ ensures compliance with the relay power constraint.
\end{theorem}
Refer to the Appendix for the proof.
Since our principal goal is to investigate the effect of correlation in a
single user parallel relay network, the rest of the paper will focus
only on the single-user case.
\section{Does Correlation Help?}
\label{Section:help_hurt}
In this section, we  address the following
important questions:
\begin{enumerate}
\item{Does correlation help when the relays know the covariance matrix?}
\item{Does correlation hurt when the relays are unaware of it?}
\end{enumerate}
To answer the first question we compare the following two scenarios.
\begin{itemize}
\item{The noise is \emph{correlated} and relays are \emph{aware} of the correlation. (Scheme-11)}
\item{The noise is uncorrelated. (Scheme-00)}
\end{itemize}
To answer the second question we compare the following two scenarios.
\begin{itemize}
\item{The noise is \emph{correlated} and relays are \emph{unaware} of the correlation. (Scheme-10)}
\item{The noise is uncorrelated. (Scheme-00)}
\end{itemize}

If Scheme-11 outperforms Scheme-00 then we can say that correlation helps if the relays are aware of it. Similarly, if Scheme-00 outperforms Scheme-10 then we can say that correlation hurts when the relays are unaware of it.
\subsection{Benchmark Schemes}
\subsubsection{Relays with uncorrelated noise: Scheme-00}
This setup has been commonly studied in literature
\cite{larsson_hong, maric_yates, mac_bc_dual, Yi2007}. The system model consists of a two hop parallel relay
network with independent noise at the relays. The noise covariance
matrix, which is diagonal, is given by $\bf K \odot I$. For a sum
power constraint at the relays, the optimum relay amplification
vector is found in \cite{maric_yates, larsson_hong}. The result can
also be obtained from Theorem 1 by replacing $\bf K$ with $\bf K
\odot I$. We thus obtain
\begin{equation}
{\bf d}_{00}=\kappa \left[({\bf GKG^{\dagger}}P_{R}+{\bf
ff^{\dagger}}P+{\bf K})\odot {\bf I}\right]^{-1}({\bf f^{\dagger}
\odot g^{\dagger}})^{T}. \label{opt_gain_uncorrelated}
\end{equation}
The SNR achieved at the destination with the optimal relay gain ${\bf d}_{00}$ is given by
\begin{equation}\mbox{SNR}_{00}=PP_{R}{\bf (f \odot g)^{\dagger}}\left[({\bf
GKG^{\dagger}}P_{R}+{\bf ff^{\dagger}}P+{\bf K})\odot {\bf
I}\right]^{-1}{\bf(f \odot g)}.\label{Eq:SNR_Scheme00}\end{equation}

A special case of Scheme-00 is where the relay noise terms are independent and identically distributed. Since we require the trace of the noise
covariance matrix to be equal in all the schemes, the covariance matrix for this scheme will be
\[{\bf K_{\mbox{iid}}}=\frac{\mbox{Tr({\bf K})}}{N}{\bf I}.\] We denote this model as Scheme-iid indicating that the relay noise terms are i.i.d. Again, the optimal relay
amplification vector for this model can be obtained from Theorem 1
by replacing $\bf K$ with $\frac{\mbox{Tr({\bf K})}}{N}{\bf I}$.
Thus we have
\begin{equation}
{\bf d}_{\mbox{iid}}=\kappa \left[\left({\bf GG^{\dagger}}\frac{\mbox{Tr({\bf
K})}}{N}P_{R}+{\bf ff^{\dagger}}P+{\frac{\mbox{Tr({\bf K})}}{N}}{\bf
I}\right)\odot {\bf I}\right]^{-1}({\bf f^{\dagger} \odot
g^{\dagger}})^{T}. \label{opt_gain_uncorrelated_iid}
\end{equation}
The corresponding SNR is given by
\begin{equation}
\mbox{SNR}_{\mbox{iid}}=PP_{R}{\bf (f \odot
g)^{\dagger}}\left[\left({\bf GG^{\dagger}}\frac{\mbox{Tr({\bf
K})}}{N}P_{R}+{\bf ff^{\dagger}}P+{\frac{\mbox{Tr({\bf K})}}{N}}{\bf
I}\right)\odot {\bf I}\right]^{-1}{\bf(f \odot g)}.
\end{equation}

%
%

%
%
%

\subsubsection{Relays with no correlation knowledge: Scheme-10}
In this scheme, correlation is induced between the relay noise
terms. However the marginals remain the same as in Scheme-00,
$n_{r_{i}} \sim \mathcal{CN}(0, {\bf K}_{ii})$; i.e. noise at relay
$R_i$ is AWGN with variance ${\bf K}_{ii}$. Note that ${\bf K}_{ij}$
represents the correlation between the noise terms $n_{R_i}$ and
$n_{R_j}$ for $i \neq j$, and in general may not be equal to zero.
Since the relays do not utilize the correlation structure, the relay
amplification vector is the same as in the case of Scheme-00, i.e.,
\[{\bf d}_{10}={\bf d}_{00}.\]
%
The SNR achieved with the relay gain ${\bf D}_{10}= \mbox{diag}({\bf d}_{10})$ is
given by
\begin{equation}
\mbox{SNR}_{10}=\frac{\left|{\bf (f \odot g)^{\dagger}}{\bf
D}_{10}^{-1}{\bf (f \odot g)}\right|^2PP_R}{{\bf (f \odot
g)^{\dagger}}{\bf D}_{10}^{-1}{\bf A}{\bf D}_{10}^{-1}(\bf f \odot
g)}.\label{Eq:SNR_Scheme10}
\end{equation}
As there is no correlation knowledge at the relays, the relay operation
here involves only co-phasing of the input signal and does not involve noise
cancelation. It is straightforward to see that Scheme-11 will
perform better than Scheme-10 as it exploits the correlation structure in designing the relay amplification factors.
Rewriting the results from Theorem \ref{THEOREM:SINGLE USER MAX RATE}, the optimal relay amplification vector that exploits full correlation knowledge
is\begin{equation}
{\bf d}_{11}=\kappa\left( \left[({\bf GKG^{\dagger}}P_{R}+{\bf
ff^{\dagger}}P+{\bf K})\right]^{-1}({\bf f
\odot g})\right)^{*}.
\end{equation}
and the SNR achieved with $\bf d_{11}$ is given by
\begin{equation}
\mbox{SNR}_{11}=PP_{R}{\bf (f \odot g)^{\dagger}}\left({\bf
GKG^{\dagger}}P_{R}+{\bf ff^{\dagger}}P+{\bf K}\right)^{-1}{\bf(f \odot g)}.\\\end{equation}
It is not clear whether Scheme-10 is also inferior
to Scheme-00 and Scheme-iid. It is also not clear which is better among Scheme-00 and Scheme-iid.
Answering these questions will provide useful insights into the impact of correlation in multi-hop AF relay
networks. We address this in the rest of this section.
\subsection{Asymptotic analysis}
We compare the schemes for the two extreme cases of $P_R \rightarrow \infty$ and $P \rightarrow \infty$.
\subsubsection{Relay Power $P_{R} \rightarrow \infty$} From the
property of the optimal network at high $P_{R}$, we have from
(\ref{mrc_bound})
\begin{equation}
\mbox{SNR}_{11}={\bf f^{\dagger}K^{-1}f}P.\label{MRC_COR}
\end{equation}
At very high $P_{R}$, ${\bf D_{10}}={\bf G(K \odot I)G^{\dagger}}
P_R$ and ${\bf A}={\bf GK G^{\dagger}} P_R$ . Substituting these in
(\ref{Eq:SNR_Scheme10}), we obtain
\begin{eqnarray}
\mbox{SNR}_{10}&=&\frac{\left|{\bf (f \odot g)}^{\dagger} ({\bf G(K
\odot I)G^{\dagger}} P_R)^{-1}{\bf (f \odot g)}\right|^2PP_R}{{\bf
(f \odot g)}^{\dagger}({\bf G(K \odot I)G^{\dagger}} P_R)^{-1}({\bf
GK G^{\dagger}} P_R)({\bf G(K \odot I)G^{\dagger}}
P_R)^{-1}{\bf (f \odot g)}} \\
&=&\frac{{\bf (f^{\dagger}(K \odot I)^{-1}f)^2}P}{{\bf f^{\dagger}(K
\odot I)^{-1}K(K \odot I)^{-1} f}}.\label{MRC_UNCOR}
\end{eqnarray}
It is evident from (\ref{Eq:SNR_Scheme00}) that the SNR of
Scheme-11 at very high relay power is
\begin{eqnarray}
\mbox{SNR}_{00}&=&P{\bf f}^{\dagger} {\bf( K \odot I) ^{-1}}{\bf f}.
\label{MRC_NOCOR}
\end{eqnarray}
Similarly the SNR of Scheme-iid is readily obtained as
\begin{eqnarray}
\mbox{SNR}_{\mbox{iid}}&=&\frac{{\bf f}^{\dagger} {\bf
f}P}{\mbox{Trace}({\bf K})/N}. \label{MRC_EQ}
\end{eqnarray}
\subsubsection*{Remark} Except for Scheme-10, all the relay schemes turn out to be a point to point SIMO channel.
It must be noted that the noise at the multiple antenna receiver
is correlated and the noise covariance matrix is the same as the relay noise covariance matrix of the original network.
For Scheme-10, the multiple antenna receiver is unaware of the noise correlation and
assumes the noise covariance matrix to be $\bf K\odot I$ instead of
$\bf K$.

To answer the question whether correlation hurts when there exists
no knowledge of it, let us consider the difference
$\mbox{SNR}_{10}-\mbox{SNR}_{00}$. For the simple case of $N=2$ and
for real channels, we have
\begin{equation}
\mbox{SNR}_{10}-\mbox{SNR}_{00}=
\frac{-2{f}_{1}{f}_{2}{\bf K}_{12}({ f}_{1}^2{\bf K}_{11}+{ f}_{2}^2{\bf K}_{22})}{{f}_{1}^2{\bf K_{11}}+{f}_{2}^2{\bf K}_{22}+2{f}_1{f}_2{\bf K}_{12}}
\end{equation}
where $ {\bf f}=\left[f_1~f_2\right]^{T}$. The difference can be either positive or negative depending on the
term $-2{f}_{1}{f}_{2}{\bf K}_{12}$. For example, when the signals are
positively correlated (${f}_{1}{ f}_{2}\geq 0$) and the noise components
are negatively correlated ${\bf K}_{12}\leq 0$, then correlation helps as
part of noise gets canceled. Similarly when both the signal and
noise components are correlated in the same direction (positive or
negative) then correlation hurts as it increases the noise power.
Therefore the overall effect of correlation can only be determined
from the average behavior.

\begin{theorem}
\label{Theorem:snr_10>snr_00: highpower}
At high relay power $P_R$, Scheme-10 outperforms Scheme-00 in terms of both average SNR and average rate. That is \[
\mathbb{E}[\mbox{SNR}_{10}]\geq \mathbb{E}[\mbox{SNR}_{00} ] ~~\mbox{and}~~\mathbb{E}[\mbox{R}_{10}]\geq \mathbb{E}[\mbox{R}_{00}]\]where $\mbox{R}_{10}=\log(1+\mbox{SNR}_{10})$ and $\mbox{R}_{00}=\log(1+\mbox{SNR}_{00})$.
\end{theorem}
The proof is presented in the Appendix.

\emph{The above result is significant as it
suggests that, in the average sense, correlation does not hurt even
if correlation knowledge is not available.} To determine the relationship for the rest of the cases, we take the
expectation of SNR in (\ref{MRC_COR}), (\ref{MRC_NOCOR}), and
(\ref{MRC_EQ}):
\begin{eqnarray}
\mathbb{E}[\mbox{SNR}_{11}]&=&P\sum_{i=1}^{N}\mathbb{E}[| {f}_{i}|^2](1/\lambda_{i})=P\sum_{i=1}^{N}1/\lambda_{i}\\
\mathbb{E}[\mbox{SNR}_{00}]&=&P\sum_{i=1}^{N}\mathbb{E}[|{ f}_{i}|^2](1/{\bf K}_{ii})=P\sum_{i=1}^{N}1/{\bf K}_{ii}\\
\mathbb{E}[\mbox{SNR}_{\mbox{iid}}]&=&\frac{P\sum_{i=1}^{N}\mathbb{E}[|{f}_{i}|^2]}{\sum_{i=1}^{N}{\bf K}_{ii}/N}=
\frac{NP}{\sum_{i=1}^{N} {\bf K}_{ii}/N}
\end{eqnarray}
where $\mathbb{E}[|{f}_{i}|^2]=1$ while $\lambda_{i}$ and ${\bf K}_{ii}$ are the $i^{th}$ eigen-value and
$i^{th}$ main-diagonal term of ${\bf K}$ respectively.
$\mathbb{E}[\mbox{SNR}_{11}]$ is greater than
$\mathbb{E}[\mbox{SNR}_{00}]$ follows from the reason that the
ordered vector containing the eigen-values of ${\bf K}$ majorizes
the main diagonal of ${\bf K}$, i.e. ${\bf \lambda \succeq d}$. We
also have $\mathbb{E}[\mbox{SNR}_{00}] \geq
\mathbb{E}[\mbox{SNR}_{\mbox{iid}}]$ due to the harmonic-arithmetic
mean inequality. Therefore we obtain the following relationship:
\begin{eqnarray}\mathbb{E}[\mbox{SNR}_{11}] \geq \mathbb{E}[\mbox{SNR}_{10}]
\geq \mathbb{E}[\mbox{SNR}_{00}] \geq
\mathbb{E}[\mbox{SNR}_{\mbox{iid}}]\label{total_relation}\end{eqnarray}

Thus in the high relay power regime, correlation does not hurt even
when the system is ignorant of the underlying noise correlation
structure. Importantly, it is also clear that there is a performance
gain when the correlation structure is exploited.

\subsubsection{Source Power $P \rightarrow \infty$}
 At very high source power, the optimal relay network (Scheme-11)
is equivalent to a MISO system. From (\ref{mrt_bound}), we have
\begin{eqnarray}
\mbox{SNR}_{11}=P_{R}{\bf g}^{\dagger} {\bf g}. \label{MRT}
\end{eqnarray}
Similarly, it can be shown that the rest of the schemes also achieve
the SNR in (\ref{MRT}) at high $P$.

\begin{equation}
\mbox{SNR}_{10}=\mbox{SNR}_{00}=\mbox{SNR}_{\mbox{iid}}=P_{R}{\bf
g}^{\dagger} {\bf g}.
\end{equation} It can be noticed that this
scenario represents the maximal ratio transmission (MRT) where the
relays act as a multiple antenna transmitter. This is intuitive as
the noise at the relays is negligible compared to the source power
and therefore the effect of the noise covariance matrix is
non-existent. We now verify the results with numerical analysis in
the following subsection.
\subsection{Numerical Results}
\begin{figure}[t]
\center
\centerline{\psfig{figure=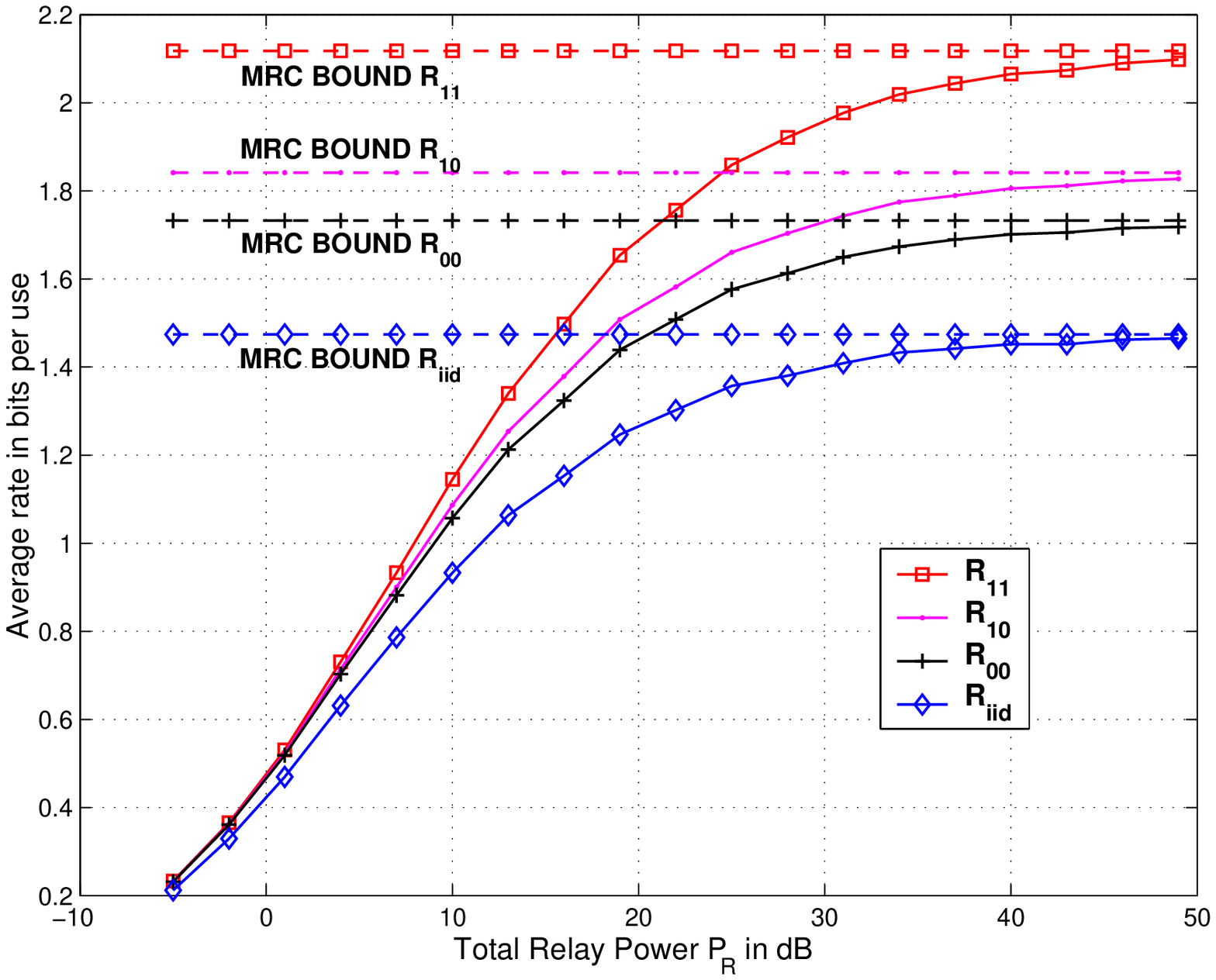,width=4.2in,height=3.2in}}
\caption{\textsf{Average rate of a single user system as a function
of total relay power $P_R$ for $P=P_{I}=10$.}} \label{relay_power}
\end{figure}

\begin{figure}[t]
\center
\centerline{\psfig{figure=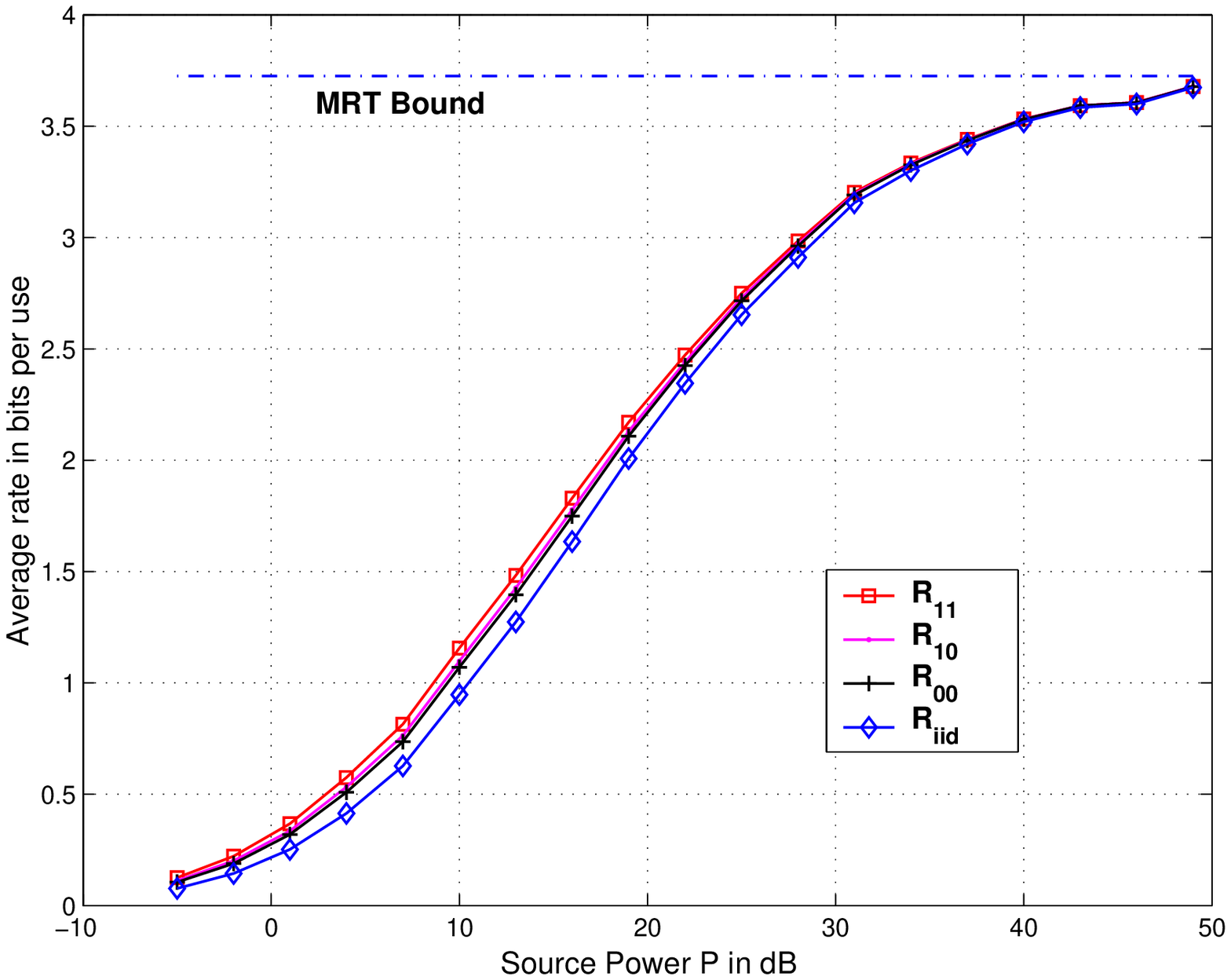,width=4.2in,height=3.2in}}
\caption{\textsf{Average rate of a single user system as a function
of source power $P$ for $P_{R}=P_{I}=10$.}} \label{source_power}
\end{figure}
 We consider a common-interference based model to generate
the relay noise covariance matrix. Each relay in addition to its
thermal noise observes faded versions of common interference signals.
The effective relay noise including both the interference and the local noise is given by
\begin{equation} {\bf n_{eff}}=\sum_{k=1}^{Q}{\bf h_{k}} i_{k}+{\bf
n_{R}}\end{equation} where $i_k$ is the signal transmitted by the $k^{th}$ interferer and ${\bf h_{k}}$ is the channel between the
relays and the $k^{th}$ interferer. The covariance matrix is given
by
\begin{equation} {\bf K}=\mathbb{E}[{\bf
n_{eff}n_{eff}^{\dagger}}]=\sum_{k=1}^{Q}{\bf
h_{k}h_{k}^{\dagger}}P_{I_k}+{\bf I}.\end{equation}

The total interference power is $P_{I}=\sum_{k=1}^{Q}P_{I_k}$. Fig.
\ref{relay_power} plots the average rate per channel-use as a
function of the total relay power for $P=P_{I}=10$ for the case of two relays and one interferer. The main
observations are
\begin{enumerate}
\item{The performance order of the schemes in (\ref{total_relation}) which was obtained at high $P_{R}$ is valid at all values of $P_{R}$.}
\item{The difference ($R_{11}-R_{10}$) which indicates the benefits of learning correlation increases with relay
power.} With increase in relay power, the schemes diverge in
performance, which is an indication that correlation impacts more at
high relay power.
\end{enumerate}

\begin{figure}[t]
\center \centerline{\psfig{figure=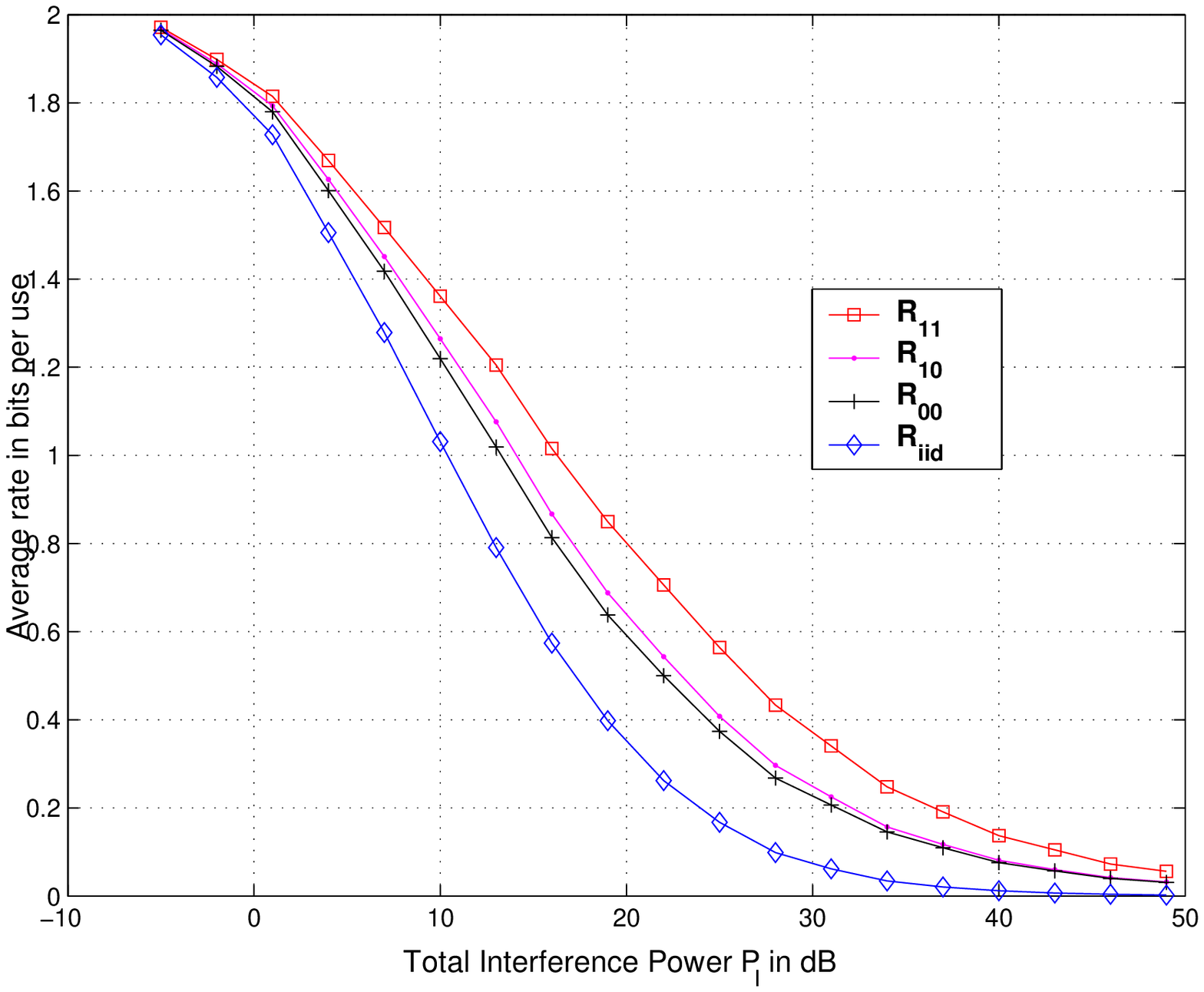,width=4in,height=3in}}
\caption{\textsf{Average rate of a single user system as a function
of interference power $P_I$ for $P=10$ and $P_{R}=100$.}}
\label{inter1}
\end{figure}

In Fig. \ref{source_power}, we plot the average rate as a function
of source transmit power for $P_R=P_I=10$. Clearly, with increase in
source power, the schemes achieve the MRT bound. Therefore the
effect of correlation is less pronounced with increasing source
power. Fig. \ref{inter1} shows the average rate as a function of the
interference power for one interfering node and two relays. As one
can expect, the average rate decreases with $P_{I}$. However
$R_{11}$ does not reach zero even at infinite interference power.
This can be explained through the following: With one interfering
source and $N$ relays, the average SNR for Scheme-11 at very high
$P_{R}$ is given by
\begin{equation} \mathbb{E}[\mbox{SNR}_{11}]=(N-1)+\mathbb{E}\left[\frac{1}{1+P_{I}\parallel
{\bf h_1}\parallel^2}\right].\end{equation} This suggests that the
average SNR is at least $N-1$ irrespective of the interference
power. However, the average SNR decreases when the number of
interfering sources increases. This is due to loss in degrees of
freedom due to increase in the range space of interference.

\begin{figure}
\center
\centerline{\psfig{figure=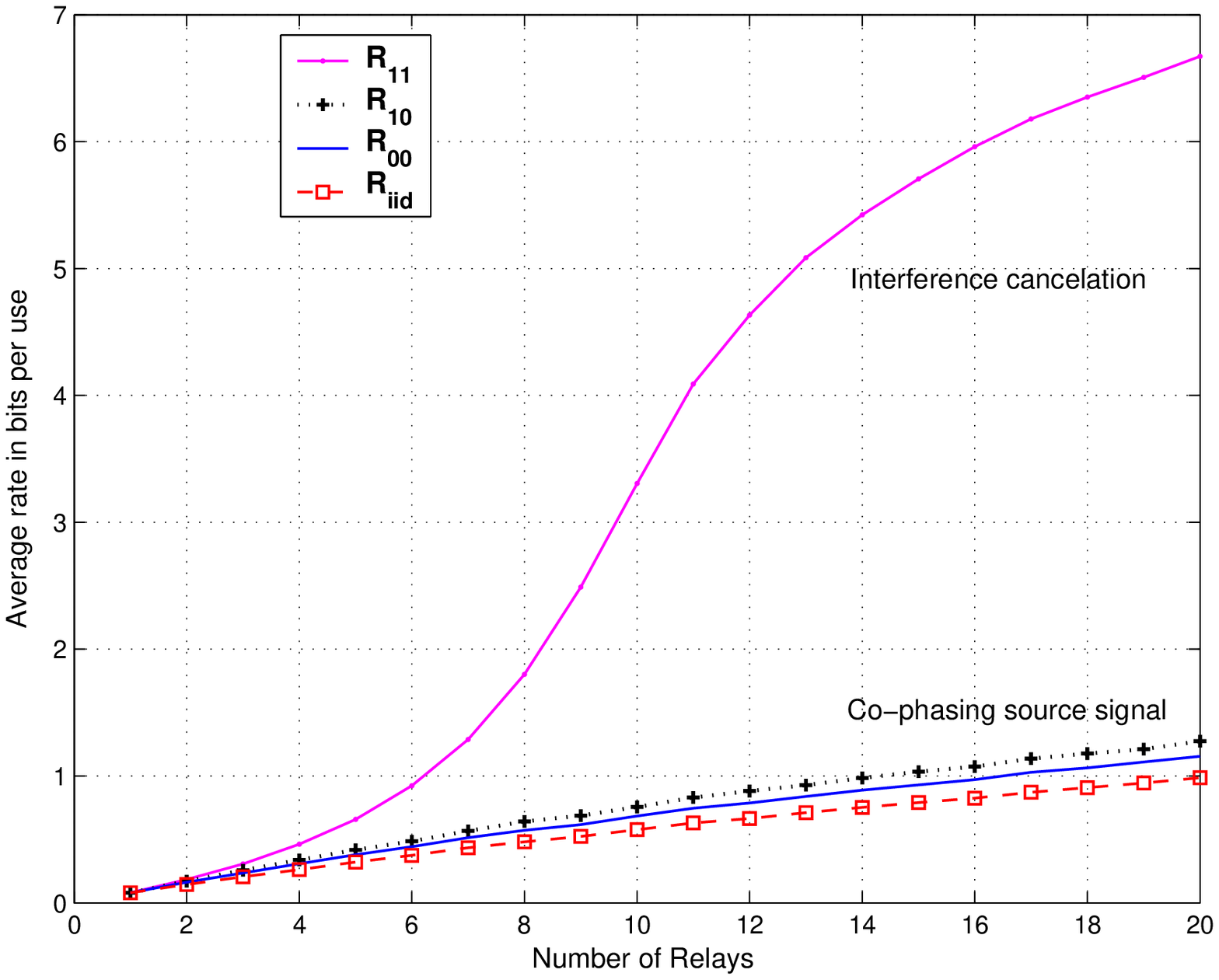,width=4in,height=3in}}
\caption{\textsf{Average rate of a single user system as a function
of the number of relays with 9 interferers of total power
$P_{I}=200$ (23 dB), and $P=10$.}} \label{relay_number}
\end{figure}

Fig. \ref{relay_number} shows the average sum rate versus the number
of relays. The covariance matrix is generated with the help of 9
interferers with total power $P_I=200$. It can be noticed that the
sum rate increases at a greater rate when the number of relays is
greater than the number of interferers. As $P_{I}\rightarrow
\infty$, $R_{11} \rightarrow 0$ when the number of relays is less
than the number of interferers ($N \leq Q$). However for ($N > Q$),
$R_{11}$ does not vanish even when $P_{I}\rightarrow \infty$.

\subsubsection*{Remark}
As we discussed earlier, the relay design problem is a tradeoff
between canceling the interference and maximizing the signal power.
At very high interference power, it is important to cancel the
interference. At high $P_{R}$ since the network behaves as a SIMO
system with $N$ antennas at the receiver, up to $(N-1)$ interfering
sources can be rejected. When the number of interferers is more than
the total number of relays, interference cannot be completely
nulled.

\subsection{Impact of channel knowledge at the relays}
Throughout this paper, we assume that the relays have perfect
knowledge of all the channels in the network. It may seem that
complete channel knowledge is required at the relays for the
relationship in (\ref{total_relation}) to hold. In Scheme-11, since
the relays know the noise covariance matrix, they must also know the
channels between the relays and interferers (when the number of
interferers is less than the number of relays). This information is
not available at the relays for Scheme-10. Still, it performs better
than Scheme-00 where there is no noise correlation. In fact, it can
be shown that Scheme-10 outperforms Scheme-00 even when there is
absolutely no channel knowledge at the relays including the
source-relay and relay-destination channels. This is also true when
the relays only have local channel information. That is, relay
$R_{i}$ knows only its forward and backward channels ${\bf f}(i)$
and ${\bf g}(i)$. The following theorem states these results:

\begin{theorem}\label{Theorem: no_csi}
The average achievable rate for a relay network is lesser with relay noise covariance matrix $\bf K \odot I$
than with any general $\bf K$, where only local channel knowledge is available at the relays. In other words
\begin{equation}
\mathbb{E}[\log(1+\mbox{SNR}_{10})] \geq \mathbb{E}[\log(1+\mbox{SNR}_{00})].
\end{equation}
Further, when there is  absolutely no CSI available at the relays, Scheme-10 outperforms Scheme-00 in terms of the average SNR, i.e. $\mathbb{E}[\mbox{SNR}_{10}] \geq \mathbb{E}[\mbox{SNR}_{00}]$.
\end{theorem}
\begin{proof}Refer to the Appendix.
\end{proof}
It is now clear that relay noise correlation is always \emph{helpful} regardless of the channel and correlation knowledge at the relays.
For maximizing signal power channel knowledge is essential while correlation knowledge is required to minimize interference power.
With perfect channel knowledge, increasing the number of relays (while keeping the total relay power fixed) is helpful as signal power is increased due to coherent combining.

\section{Three-hop parallel relay network}
\label{Section:3hops}
\begin{figure}
\center
\input{three_hop.pstex_t}
\caption{\textsf{Three-hop parallel relay
network}}\label{three_hop_single_user}
\end{figure}
In this section, we consider another application of the relay
network with correlated noise. Consider a three-hop relay network as
shown in Fig. \ref{three_hop_single_user}. This is an example of a
multi-stage relay network. There are $N$ and $M$ relays in the first
and second stages respectively. Let the power expended by the
source, the first stage relays, the second stage relays be $P_0$,
$P_1$ and $P_2$ respectively.  We denote this network with the shorthand notation $\mathfrak{R}_{3}( P_0,{\bf f},{\bf d_1}, P_1, {\bf
H},{\bf d_2}, P_2, {\bf g} )$. In the first slot, the received
signal at the first stage relays is given by
\begin{equation}
{\bf r_{1}=f}x+{\bf n_{1}}.
\end{equation}
The relay transmitted signal is given by $\bf
{t_{r_{1}}=D_{1}r_{1}}$, where $\bf D_1$ satisfies the power
constraint of the first stage relays, i.e.
$\mbox{Tr}({\bf{D_1D_1^{\dagger}}({\bf ff^{\dagger}}P+{\bf I})})=P_1$. Here
$\bf D_1=\mbox{diag}(d_1)$. In the second time slot, the relays at
the second stage receive
\begin{equation}
\bf{r_2=HD_1r_1+n_{2}}= \bf {HD_1f}x+{\bf HD_1n_{1}+n_2}.
\end{equation}
In the third slot, the signal transmitted by the second stage relays
is $\bf t_{r_2}=D_{2}r_2$. The power constraint at the second stage
relays is given by $\mathbb{E}\left[\parallel\bf
t_{r_2}\parallel^2\right]=P_2$, which results in
\begin{eqnarray}\mbox{Tr}\left({\bf D_2D_2^{\dagger}}({\bf HD_1ff^{\dagger}D_1^{\dagger}H^{\dagger}}P_0+{\bf
HD_1D_1^{\dagger}H^{\dagger}+I})\right)=P_2.\end{eqnarray} Notice
that the choice of $\bf D_1$ affects the power constraint at $\bf
D_2$. The received signal at the destination is given by
\begin{eqnarray}
y={\bf g^{T}t_{r_2}}&=&{\bf g^{T}D_2}\left({\bf HD_1}({\bf f}x+{\bf
n_1})+{\bf n_2}\right)+n\\
&=&{\bf g^{T}D_2HD_1f}x+{\bf g^{T}D_2HD_1n_1}+{\bf g^{T}D_2n_2}+n.
\label{input_output_three_hops}
\end{eqnarray}
Note that both the relay stages operate at their maximum
power. For any feasible $\bf D_2$ and $\bf D_1$, the SNR is
defined as
\begin{equation}\mbox{SNR}({\bf D_1}, {\bf D_2} )=\frac{\left|{\bf
g^{T}D_2HD_1f}\right|^2P_0}{{\parallel\bf
g^{T}D_2HD_1\parallel^2}+\parallel {\bf g^{T}D_2}\parallel^2+1}.
\label{EQUATION_THREE_HOP_SNR}\end{equation} The relay design
problem is stated below as
\begin{equation}
\max _{{\bf D1, D_2}}\mbox{SNR}({\bf D_1, D_2}),
\end{equation}
\begin{equation}
\mbox{s.t.}~~~\mbox{Tr}({\bf{D_1D_1^{\dagger}}({\bf
ff^{\dagger}}P+{\bf I})})=P_1, \label{EQUATION_CONSTRAINT_FIRST_HOP}
\end{equation}
\begin{equation}
\mbox{and}~~~~\mbox{Tr}\left({\bf D_2D_2^{\dagger}}({\bf
HD_1ff^{\dagger}D_1^{\dagger}H_2^{\dagger}}P_0+{\bf
HD_1D_1^{\dagger}H^{\dagger}+I})\right)=P_2.
\label{EQUATION_CONSTRAINT_SECOND_HOP}
\end{equation}

To solve this problem directly can be difficult as the constraints are
interdependent. We therefore approach the problem differently. Now
consider any feasible $\bf D_1$ for the first stage relays. Once the
relay operation for the first stage relays is fixed, the network
reduces to a two hop parallel relay network with correlated noise
$\mathfrak{R}_{2}(P_{0}, {\bf f_{\mbox{new}}}, {\bf D_{2}}, P_{2},
{\bf g})$
 where
\begin{eqnarray}
{\bf f_{\mbox{new}}=HD_1f}.
\end{eqnarray}
By comparing (\ref{input_output_three_hops}) and
(\ref{received_signal}), relay noise in the reduced two hop
network can be found as
\begin{equation}
{\bf n_R=HD_{1}n_{1}+n_{2}}. \label{effective noise}
\end{equation}
Note that the noise in (\ref{effective noise}) is correlated and the
corresponding covariance matrix is given by
\begin{equation}
{\bf K=HD_1D_1^{\dagger}H^{\dagger}+I}.
\end{equation}
Now that the network is a dual-hop parallel relay network, we can
utilize the result in Theorem \ref{THEOREM:SINGLE USER MAX RATE} to
solve the relay optimization for the second stage relays. We have
from (\ref{corr_amp_vector})
\begin{eqnarray}
{\bf d_2^{\circ}}({\bf D_{1}})&=&\kappa \left(\left[{\bf
GKG^{\dagger}}P_2+({\bf f_{\mbox{new}}f_{\mbox{new}}^{\dagger}\odot
I})P_0+{\bf K \odot
I}\right]^{-1}({\bf f_{\mbox{new}} \odot g})\right)^{*}. 
\end{eqnarray}

We therefore are able to determine the optimal design for the second
stage relays for all feasible relay gain vectors for the first
stage. Similarly if we can find optimal $\bf D_{1}$ for all possible
$\bf D_2$ we can then resort to iterating between the solutions.
However the problem is that the second stage relays' power constraint is affected by $\bf D_1$.
Therefore fixing $\bf D_2$ and finding the optimal $\bf D_1$ can
violate the power constraint for the second stage. One way to solve
this problem is to use the reciprocity property of AF relay networks
recently proved in \cite{mac_bc_dual}. We state the theorem from
\cite{mac_bc_dual} in the following.
\begin{theorem}{\cite{mac_bc_dual}} The capacity of
a three-hop relay network is unchanged when the role of the
transmitter and the receiver is switched while maintaining the same
transmit power at each hop for both the original and reciprocal
channels. This is equivalent to
\begin{equation}\mathfrak{R}_{3}( P_0,{\bf f},{\bf d_1}, P_1, {\bf H},{\bf
d_2}, P_2, {\bf g})=\mathfrak{R}_{3}( P_2,{\bf g},\kappa_2{\bf d_2},
P_1, {\bf H^{T}},\kappa_{1}{\bf d_1}, P_0, {\bf f} ),\end{equation}
where the constants $\kappa_1$ and $\kappa_2$ ensures power
constraint compliance in the reciprocal network. (Note that we allow
some abuse of notation as the above statement implies that the
capacity of the two networks are the same.)
\end{theorem}

Using the above result, one can reverse the direction of the
communication in the relay network. For the reciprocal network, the
gain of the first stage relays is related to the gain of the second
stage relays through a constant multiplying factor which can be
obtained from the power constraint as follows:
\begin{equation}
\kappa_2^2=\frac{P_1}{\mbox{Tr}\left({\bf D_2D_2^{\dagger}}({\bf
gg^{\dagger}}P_2+{\bf I})\right)}
\end{equation}
Similarly $\kappa_1$ is found as
\begin{equation}
\kappa_1^2=\frac{P_0}{\mbox{Tr}\left({\bf
D_1D_1^{\dagger}}(\kappa_2^2{\bf
H^{T}D_2gg^{\dagger}D_2^{\dagger}H^{*}}P_2+\kappa_2^2{\bf
H^{T}D_2D_2^{\dagger}H^{*}+I})\right)}.\end{equation}

As discussed earlier, the reciprocal network can be reduced to a two
hop network and the relay gain for the second stage can be
optimized. Using Theorem \ref{THEOREM:SINGLE USER MAX RATE}, we can
find the optimal gain for the second stage of the reciprocal network
(which is proportional to the first stage relay of the original
network) for every feasible $\kappa_2{\bf D_2}$. The procedure then
will be to find the optimal relay gain for the second stage having
the first stage gain fixed. The next step is to find the reciprocal
network and optimize the second stage, and iterate this procedure till
convergence is achieved. The algorithm is listed in Table \ref{TABLE
ALGORITHM}.
\begin{table}
\caption{Algorithm: three-hop Relay Network Optimization}
\begin{center}
\begin{enumerate}
\item Choose any ${\bf D_1}$ that satisfies the first stage power constraint.
\item Reduce the network to a two hop network with noise correlation.
\item Find the optimal ${\bf D_2^{\circ}(D_{1})}$.
\item Begin Loop:
\item Determine the reciprocal network.
\item Optimize the second stage relays.
\item Repeat until convergence.
\item End Loop
\end{enumerate}
\end{center}
\label{TABLE ALGORITHM}
\end{table}

\subsubsection*{Remark}
Notice that the capacity of the network increases with every
iteration as each iteration involves optimization. This suggests
that the algorithm will eventually converge. Although it is not known if the algorithm
guarantees a global optimum  solution, based on extensive numerical simulations, we conjecture that
the algorithm outputs the global optima whenever there is
convergence.
\begin{figure}
\center
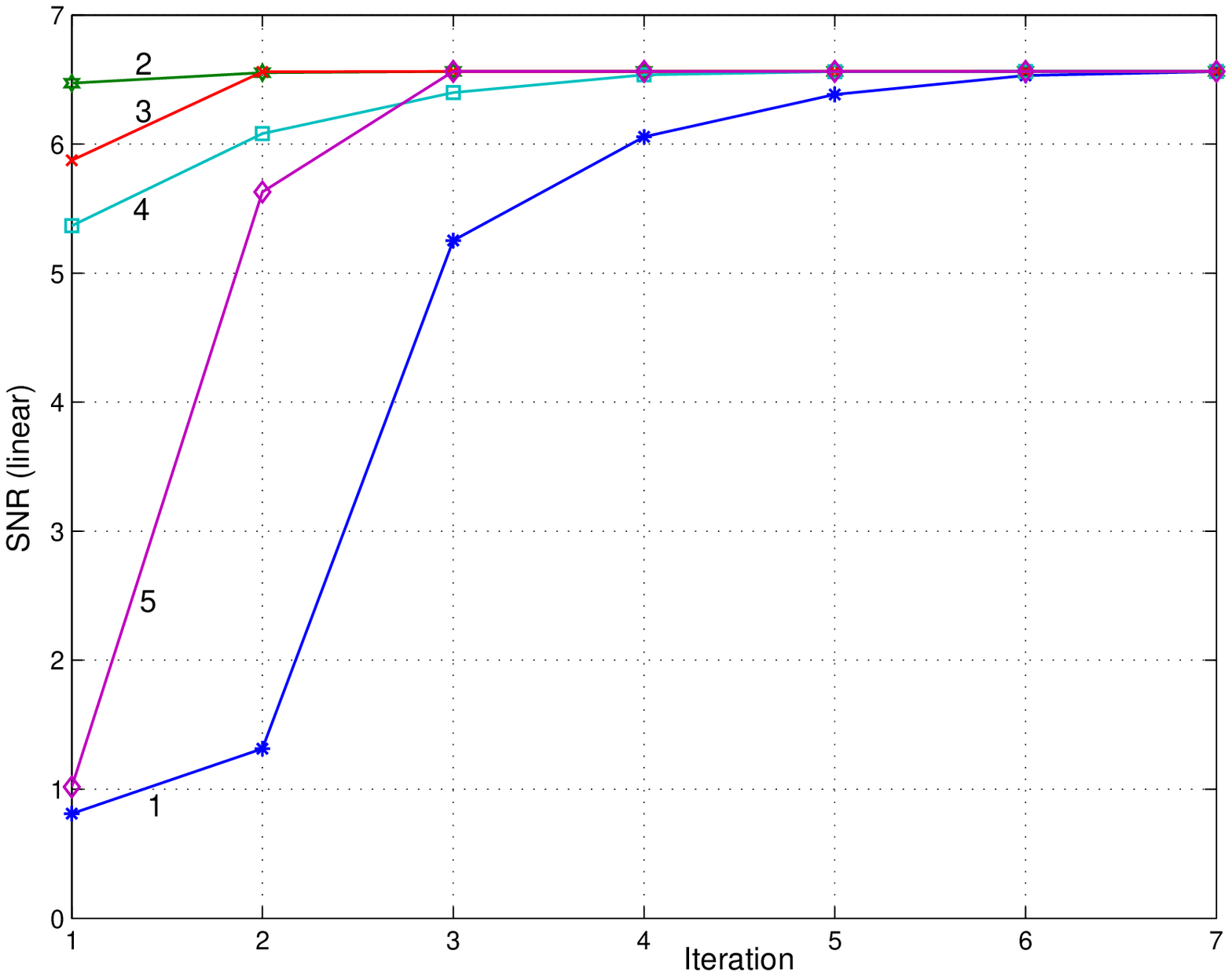
\caption{\textsf{Pictorial representation of the relay optimization
algorithm.}}\label{algorithm_pic}
\end{figure}

\begin{figure}[t]
\center
\centerline{\psfig{figure=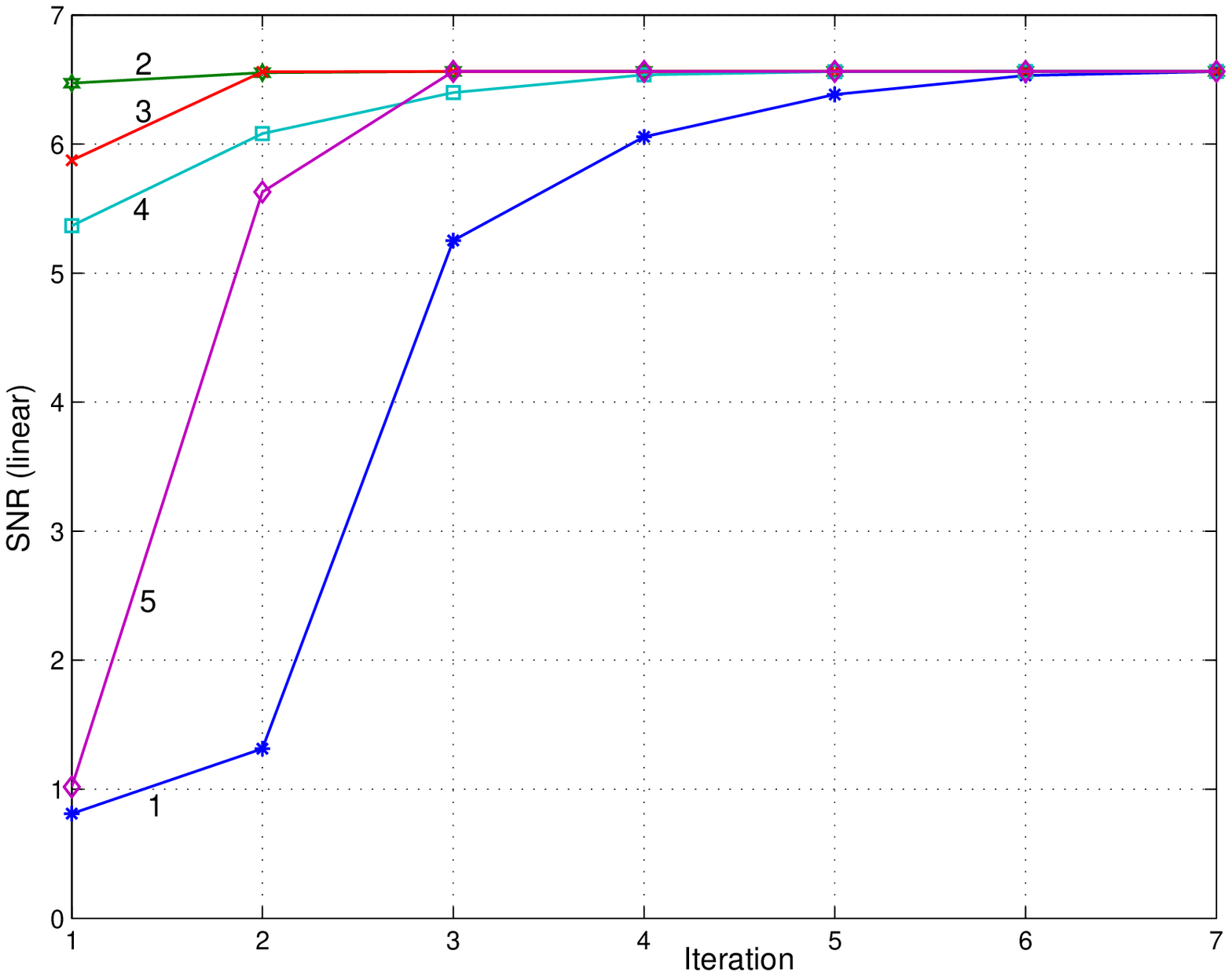,width=3.8in,height=2.8in}}
\caption{\textsf{Convergence of the algorithm for different initial inputs (${\bf d_1}$).}} \label{algorithm}
\end{figure}

Fig. \ref{algorithm} shows the convergence of the algorithm for a
variety of initial relay amplification factors. In the simulation,
the channel matrices are
\[{\bf f}=[1 ~6],~ {\bf g}=[4 ~-3],~{\bf H}=[2~ -3; 4~ 2].\]
We use the following initial inputs for the first stage relay gains
every time the algorithm begins:
\[{\bf d_1}(1)=[1~ 0],~{\bf d_1}(2)=[0 ~1], ~{\bf d_1}(3)=[-2~1], ~{\bf d_1}(4)=[2~1],~{\bf d_1}(5)=[-20,~-1].\]
The transmit power at each hop is unity ($P_0=P_1=P_2=1$). At the
start of the algorithm, the inputs will be scaled to satisfy the
power constraint. For each input, the progress of the algorithm
(measured in terms of SNR at the destination) is displayed in Fig.
\ref{algorithm}. For all these inputs, the algorithm converges to an
SNR value of 6.5638 while the first and second stage relay gains
converge to
\[{\bf \hat{d_1}}=[-0.0823~ -0.1633], ~{\bf \hat{d_2}}=[0.2533~
0.2566].\]

In the following we study the characteristics of three-hop relay
networks in the high-power regimes. The proofs are available in the
Appendix.
\subsection{Properties of $\mathfrak{R}_{3}( P_0,{\bf f},{\bf d_1}, P_1, {\bf H},{\bf d_2}, P_2, {\bf g} )$}
\label{SUBSECTION:PROPERTIES_THREE_HOP}
\begin{enumerate}
\item{$P_2 \rightarrow \infty$}:~ When there is no power constraint (or infinite power) for the last stage
relays, the three-hop network for the optimal $\bf D_2$ reduces to a two hop network with the destination consisting of multiple antennas.
This is equivalent to \begin{equation}\mathfrak{R}_{3}(P_0,{\bf
f},{\bf d_1}, P_1, {\bf H},{\bf d_2=d_2^{\circ}}, P_2=\infty, {\bf
g})=\mathfrak{R}_{2}( P_0,{\bf f},{\bf d_1}, P_1, {\bf H}
).\end{equation} It is worth noting that the noise at the multiple antenna destination of the reduced network is correlated.
\item{$P_0 \rightarrow \infty$}:~
For the case where the source power is infinite, it is clear from
the properties in Section \ref{SUBSECTION: PROPERTIES PROOF DUAL
HOP} that the noise terms at the first stage relays are negligible.
This reduces the network to a two-hop relay network with a multiple
antenna source.
\[\mathfrak{R}_{3}(
P_0=\infty,{\bf f},{\bf d_1=d_1^{\circ}}, P_1, {\bf H},{\bf d_2},
P_2, {\bf g})=\mathfrak{R}_{2}(P_1, {\bf H},{\bf d_2}, P_2, {\bf
g})\] The network optimization for the reduced network involves
joint precoder (transmit covariance matrix) and relay design which
are interdependent. We can utilize the algorithm in Table \ref{TABLE
ALGORITHM} for this optimization.

\item{$P_0, P_2\rightarrow  \infty$}:~
When the power of both the source and the last stage relays tend to
$\infty$, the relay network reduces to a point to point MIMO channel
with the constraint that the rank of the input covariance matrix is
one.\[\mathfrak{R}_{3}( P_0=\infty,{\bf f},{\bf d_1=d_1^{\circ}},
P_1, {\bf H},{\bf d_2=d_2^{\circ}}, P_2, {\bf
g})=\mathfrak{R}_{1}(P_1, {\bf H})\]
Here $\mathfrak{R}_{1}( P_1,{\bf H})$ indicates that the network is a single-hop point to point channel. Thus the three-hop network optimization problem is equivalent to SNR maximization problem in the point to point MIMO channel. The optimal strategy is beamforming along the principal eigen-vector of the channel matrix.

\item{$P_1, P_2\rightarrow  \infty$}:~
For this scenario the last two stages can be removed and the network
reduces to a point to point SIMO channel with receive combining
vector $\bf g^{T}D_2HD_1$. Optimizing one set of relays is enough to
achieve the point to point channel capacity.
\[\mathfrak{R}_{3}(
P_0,{\bf f},{\bf d_1}, P_1, {\bf H},{\bf d_2=d_2^{\circ}},
P_2=\infty, {\bf g})=\mathfrak{R}_{1}( P_0,{\bf f}).\]
$\mathfrak{R}_{1}( P_0,{\bf f})$ indicates that the network is a single-hop point to point channel.
\item{$P_0, P_1 \rightarrow  \infty$}:~
When the source and the first stage relays have very high power, the
network reduces to a point to point MISO system with transmit
precoding vector $\bf D_2HD_1f$. Like the previous case, optimizing
one set of relays suffices to optimize the whole network.
\[\mathfrak{R}_{3}(P_0=\infty,{\bf f},{\bf d_1}, P_1=\infty, {\bf H},{\bf d_2=d_2^{\circ}},
P_2, {\bf g})=\mathfrak{R}_{1}(P_2, {\bf g})\]

%
\end{enumerate}

\subsubsection*{Remark}In all the above cases, if the transmit power
at any hop is very high, noise at the receive side is negligible.
When this occurs at the first or the last hop, (for example, $P_0$
or $P_2 \rightarrow \infty$) the hop can be removed. Here any rate
that can be achieved in the original network is  achievable in the
reduced network and vice versa. For example consider the three-hop
case where the source has one antenna, and 5 relays in the first
stage. When the source power is very high ($P_0 \rightarrow \infty$) then the
first hop network can be removed, and the reduced network consists
of two hops. In the new network the source consists of 5 antennas
and transmits a one dimensional signal. It is interesting that the
two networks are equivalent.

\section{Conclusion}\label{Section:conclusion}
In this work, we considered an AF relay network wherein the relay noises are correlated which may be due to
common interference or multi-hop AF relaying. We obtained closed
form expressions for optimal rate maximizing relay gains and maximum
achievable rate when correlation knowledge is available at the
relays for both single-user and multi-source scenarios. Further we
showed that correlation does not hurt irrespective of channel and correlation knowledge at the relays.
We also showed that correlation knowledge
results in significant performance improvement. Analytical and
simulation results demonstrate significant rate enhancement when
correlation knowledge is exploited. With appropriate gains, the relays can perform
distributed interference cancelation when the number of relays is greater than the the number of interferers.
The performance improvement increases with the total relay power and the number of relays. As
there are significant benefits in learning correlation, practical
schemes to communicate the correlation structure to the relays need
to be explored.

For three-hop AF relay networks, we addressed the relay optimization
problem. By fixing relay gains for one of the stages we reduced the
network to a two hop network with noise correlation. Combining the
result we obtained for two hop networks with correlated noise and a multi-hop duality result from \cite{mac_bc_dual}, we
propose an iterative algorithm that increases capacity for every
iteration. This algorithm can be utilized for joint optimization of
relay gains and precoder in multi-antenna sources.

\begin{appendix}
\subsection{Proof of Theorem \ref{THEOREM:SINGLE USER MAX RATE}}
From the destination received signal in (\ref{received_signal}), the
achievable rate is given by $R=\log_{2}(1+\mbox{SNR})$ where
\begin{equation}\mbox{SNR}=\frac{{\bf d^{T}G}{\bf ff^{\dagger}}{\bf
G^{\dagger}d^{*}}P}{{\bf
d^{T}GKG^{\dagger}d^{*}}+1}.\label{eqn:snr_singleuser}\end{equation}
It is worth noting that ${\bf d}$ should satisfy the sum power
constraint of the relays.
\begin{equation}{\bf
d^{T}}\left[\left({\bf ff^{\dagger}}P+{\bf K}\right)\odot {\bf
I}\right]{\bf d^{*}}=P_{R} \label{eqn:power_constraint}
\end{equation}
 Now we are interested in maximizing the sum rate over all possible
relay amplification vectors that satisfy the relay power constraint.
The above problem is equivalent to maximizing SNR with respect to
${\bf d}$. \[\mbox{SNR}^{\circ}=\max_{\bf d} \mbox{SNR}\]
Incorporating (\ref{eqn:power_constraint}) into the denominator of
(\ref{eqn:snr_singleuser}), we have
\begin{alignat*}{2}
\mbox{SNR}&=&\frac{{\bf d^{T}G}{\bf ff^{\dagger}}{\bf
G^{\dagger}d^{*}}PP_{R}}{{\bf d^{T}GKG^{\dagger}d^{*}}P_{R}+{\bf
d^{T}}\left[\left({\bf ff^{\dagger}}P+{\bf K}\right)\odot {\bf
I}\right]{\bf d^{*}}} \nonumber\end{alignat*}
\begin{eqnarray}
&=&\frac{{\bf d^{T}G}{\bf ff^{\dagger}}{\bf
G^{\dagger}d^{*}}PP_{R}}{{\bf d^{T}}\left[{\bf GKG^{\dagger}} P_{R}
+\left({\bf ff^{\dagger}}P+{\bf K}\right)\odot{\bf I}\right]{\bf
d^{*}}}
\end{eqnarray}
Define the matrices {\bf A} and {\bf B}: \begin{equation} {\bf
A}=\left[{\bf GKG^{\dagger}} P_{R} +\left({\bf ff^{\dagger}}P+{\bf
K}\right)\odot{\bf I}\right]\end{equation}
\begin{equation}{\bf B}={\bf G}{\bf
ff^{\dagger}}{\bf G^{\dagger}}PP_{R}\end{equation} Notice that the
matrix ${\bf A}$ is Hermitian and positive definite while $\bf B$ is
Hermitian and positive semi-definite. By Cholesky decomposition
${\bf A=LL^{\dagger}}$, we have
\begin{eqnarray}
\mbox{SNR}= \frac{{\bf d^{T}Bd^{*}}}{{\bf d^{T}LL^{\dagger}
d^{*}}}\end{eqnarray} Let ${\bf w^{T}=d^{T}L}$, we have
\begin{eqnarray}
\mbox{SNR}= \frac{{\bf w^{T}L^{-1}BL^{-\dagger}w^{*}}}{{\bf
w^{T}w^{*}}}\end{eqnarray} The relay optimization problem is
therefore
\begin{eqnarray}
\mbox{SNR}^{\circ}= \max_{{\bf w}}\frac{{\bf
w^{T}L^{-1}BL^{-\dagger}w^{*}}}{{\bf w^{T}w^{*}}}
\end{eqnarray} By utilizing
 Rayleigh Quotient we have
\begin{eqnarray}
\mbox{SNR}^{\circ}=\lambda_{\max}({\bf L^{-1}BL^{-\dagger}})
\end{eqnarray}
\begin{eqnarray}
{\bf w}^{\circ}=\left({\bf v}_{\max}({\bf L^{-1}BL^{-\dagger}})\right)^{*}
\end{eqnarray}
where $\lambda_{\max}({\bf P})$ and $v_{\max}({\bf P})$ are the
principal eigen-value and eigen-vector of the matrix ${\bf
P}$.\\Since ${\bf B=(f\odot g)(f\odot g)^{\dagger}}PP_{R}$, we have
\begin{eqnarray}
\mbox{SNR}^{\circ} &=&PP_{R}\lambda_{\max}({\bf L^{-1}(f\odot
g)(f\odot
g)^{\dagger}L^{-\dagger}})\\
&=&PP_{R}\lambda_{\max}({\bf (f\odot
g)^{\dagger}L^{-\dagger}L^{-1}(f\odot g)})\label{eqn:a}\\
&=&PP_{R}\lambda_{\max}({\bf (f\odot
g)^{\dagger}A^{-1}(f\odot g)})\label{eqn:b}\\
&=&PP_{R}{\bf (f\odot g)^{\dagger}A^{-1}(f\odot g)}
\end{eqnarray} where (\ref{eqn:a}) follows from the property that the non-zero
eigen values of $\bf AB$ are equal to those of $\bf BA$.
(\ref{eqn:b}) holds since $\bf
L^{-\dagger}L^{-1}=(LL^{\dagger})^{-1}=A^{-1}$. The optimal relay
amplification vector is given by
\begin{eqnarray}
{\bf d}^{\circ}&=&\kappa \left({\bf L^{-\dagger}}{\bf v}_{\max}({\bf L^{-1}BL^{-\dagger}})\right)^{*}\\
&=&\kappa \left({\bf L^{-\dagger}}{\bf L^{-1}} ({\bf f \odot g})\right)^{*}\\
&=&\kappa \left({\bf A^{-1}(f \odot g)}\right)^{*}.
\end{eqnarray} where $\kappa$ ensures compliance with the relay power constraint.
Hence proved. \hfill \QED

\subsection{Proof of Properties in Section \ref{SUBSECTION: PROPERTIES PROOF DUAL HOP}}
\begin{enumerate}
\item{$P_{R}\rightarrow \infty$}:~At high relay power $P_R$, for any relay amplification vector $\bf
d$, the destination noise is negligible. In that case, the
input-output relationship can be expressed as
\[y={\bf (d \odot g)^{T}f}x+{\bf (d \odot g)^{T}n_{R}}, \]
which is equivalent to a SIMO system, ${\bf y=f}x+{\bf n_{R}}$ ,
with the receive combining vector $\bf d \odot g$. Now consider the
maximum achievable SNR (corresponding to $\bf d^{\circ}$) from
(\ref{optimal_snr_dual_hop}), which is
\[\mbox{SNR}^{\circ}=PP_{R}({\bf f\odot g})^{\dagger} {\bf A^{-1}}( {\bf
f\odot g}).\] At high $P_R$, ${\bf A=(GKG^{\dagger})}P_{R}$ which
leads to
\[\mbox{SNR}^{\circ}=P({\bf f\odot g})^{\dagger} {\bf (GKG^{\dagger})^{-1}}( {\bf
f\odot g})=P{\bf f^{\dagger} K^{-1}f}.\]
At high $P_R$, the optimal relay functionality becomes
\[ {\bf d^{\circ}}=\kappa{\bf (GKG^{\dagger})^{-1}}( {\bf
f\odot g})=\kappa {\bf G^{-\dagger}K^{-1}G^{-1}Gf}=\kappa {\bf G^{-\dagger}K^{-1}f}. \]
Thus ${\bf d =\kappa G^{-\dagger}K^{-1}f}$ achieves the MRC bound when $P_{R} \rightarrow \infty$.
\item{$P\rightarrow \infty$}:~
When the source power is high, it can be observed that the relay
noise terms are negligible. The input-output relationship is then
given by \[y={\bf g^{T}\frac{(d \odot f)}{\sqrt{(d \odot
f)^{\dagger}(d \odot f)}}} \sqrt{{\bf(d \odot f)^{\dagger}(d \odot
f)}} x+n.\] which represents a MISO system with the beamforming
vector $\bf \frac{(d \odot f)}{\sqrt{(d \odot f)^{\dagger}(d \odot
f)}}$. Since ${\bf (d \odot f)^{\dagger}(d \odot f)}P=P_{R}$, the
input transmit power is $P_{R}$. For the optimal relay design $\bf
d^{\circ}$ we have ${\bf A=(ff^{\dagger} \odot I)}P$, which results
in
\begin{eqnarray*}
\mbox{SNR}^{\circ}=P_{R}({\bf f\odot g})^{\dagger} {\bf
(ff^{\dagger} \odot I)^{-1}}( {\bf f\odot g})=P_{R}{\bf g^{\dagger}
g}.
\end{eqnarray*} The system thus behaves as a multiple antenna transmitter with a single antenna receiver.\hfill \QED
\end{enumerate}
\subsection{Proof of Theorem \ref{THEOREM: MULTIUSER MAX RATE}}
From the destination received signal in (\ref{received_signal_mac}), we
can notice that due to relay amplification, the relay network is
equivalent to a scalar MAC. The sum rate is given by
\[R=\log(1+\mbox{SNR})\] where
\begin{equation}\mbox{SNR}=\frac{{\bf d^{T}G}\left(\sum_{k=1}^{L}{\bf
f_{k}f_{k}^{\dagger}}P_{k}\right){\bf G^{\dagger}d^{*}}}{{\bf
d^{T}GKGd^{*}}+1}.\label{eqn:snr_mac}\end{equation} Notice that
${\bf d}$ satisfies the sum power constraint of the relays.
\begin{equation}{\bf
d^{T}}\left[\left(\sum_{k=1}^{L}{\bf f_{k}f_{k}^{\dagger}}P_{k}+{\bf
K}\right)\odot {\bf I}\right]{\bf d^{*}}=P_{R}
\label{eqn:power_constraint_mac}
\end{equation}
Now we are interested in maximizing the sum rate over all possible
relay amplification vectors that satisfy the relay power constraint.
The above problem is equivalent to maximizing SNR with respect to
${\bf d}$. \[\mbox{SNR}^{\circ}=\max_{\bf d} \mbox{SNR}\]
Incorporating (\ref{eqn:power_constraint_mac}) into the denominator
of (\ref{eqn:snr_mac}), we have
\begin{alignat*}{2}
\mbox{SNR}&=&\frac{{\bf d^{T}G}\left(\sum_{k=1}^{L}{\bf
f_{k}f_{k}^{\dagger}}P_{k}\right) {\bf G^{\dagger}d^{*}}P_{R}}{{\bf
d^{T}GKG^{\dagger}d^{*}}P_{R}+{\bf
d^{T}}\left[\left(\sum_{k=1}^{L}{\bf f_{k}f_{k}^{\dagger}}P_{k}+{\bf
K}\right)\odot {\bf I}\right]{\bf d^{*}}} \nonumber\end{alignat*}
\begin{eqnarray}
&=&\frac{{\bf d^{T}G}\left(\sum_{k=1}^{L}{\bf
f_{k}f_{k}^{\dagger}}P_{k}\right){\bf G^{\dagger}d^{*}}P_{R}}{{\bf
d^{T}}\left[{\bf GKG^{\dagger}} P_{R} +\left(\sum_{k=1}^{L}{\bf
f_{k}f_{k}^{\dagger}}P_{k}+{\bf K}\right)\odot{\bf I}\right]{\bf
d^{*}}}
\end{eqnarray}
Define the matrices {\bf A} and {\bf B}: \begin{equation} {\bf
A}=\left[{\bf GKG^{\dagger}} P_{R} +\left(\sum_{k=1}^{L}{\bf
f_{k}f_{k}^{\dagger}}P_{k}+{\bf K}\right)\odot{\bf
I}\right]\end{equation}
\begin{equation}{\bf B}={\bf G}\left(\sum_{k=1}^{L}{\bf
f_{k}f_{k}^{\dagger}}P_{k}\right){\bf
G^{\dagger}}P_{R}\end{equation}

Then, proceeding as in the proof for Theorem 1, we have.
\begin{eqnarray}
\mbox{SNR}^{\circ}
&=&\lambda_{\max}({\bf A^{-1}B})
\end{eqnarray}
and the optimal relay amplification vector is given by
\begin{eqnarray}
{\bf d}^{\circ}=\left({\bf v}_{\max}({\bf A^{-1}B})\right)^{*}.
\end{eqnarray}
Hence proved. \hfill\QED
\end{appendix}

\subsection{Proof of Theorem \ref{Theorem:snr_10>snr_00: highpower}}
From (\ref{MRC_UNCOR}) and (\ref{MRC_NOCOR}) we have
\begin{equation}
\mbox{SNR}_{00}=a~~~~~~~~\mbox{SNR}_{10}=\frac{a^2}{a+b}
\end{equation}
where
\begin{equation}
a=\sum_{i=1}^{N} \frac{|{\bf f} (i)|^{2}}{{\bf K}_{ii}}.
\end{equation}
\begin{equation}
b=\sum_{i=1}^{N}\sum_{j=1~j\neq i}^{N}\frac{{\bf f}(i){\bf f^{*}}(j){\bf K}_{ij}}{{\bf K}_{ii}{\bf K}_{jj}}
\end{equation}
Now consider the the difference in SNR of Scheme-10 and Scheme-00, which is given by \begin{equation}
\mbox{SNR}_{10}-\mbox{SNR}_{00}=\frac{-ab}{a+b}
\end{equation}
Notice that the difference $\mbox{SNR}_{10}-\mbox{SNR}_{00}$ is
convex in the correlation term $b$ as its second derivative w.r.t.
$b$ is non-negative. We assume that the magnitude and phase of the
components of ${\bf f}$ are independent where the magnitude follows
Rayleigh distribution while the phase is uniformly distributed in
the interval $[0, 2\pi]$. Conditioned on the magnitude of the
channels, the average of $b$ over all channel phases is 0. By
applying Jensen's inequality we have
\[\mathbb{E}[\mbox{SNR}_{10}-\mbox{SNR}_{00}]\geq 0.\]
Let us now consider the difference in rate $R_{10}-R_{00}$.
\begin{equation}
R_{10}-R_{00}=\mathbb{E}\left[\log\left(\frac{ 1+\mbox{SNR}_{10}}{1+\mbox{SNR}_{00}}\right)\right]=\log\left(\frac{1+\frac{a^2}{a+b}}{1+a}\right).
\end{equation}
It can be shown that the difference $R_{10}-R_{00}$ is convex in $b$. Therefore through Jensen's inequality
\begin{equation}
\mathbb{E}[R_{10}-R_{00}]\geq 0.
\end{equation}
Hence proved. \hfill \QED
\subsection{Proof of Theorem \ref{Theorem: no_csi}}
For any constant relay gain $\bf D$ for the relays, the SNR of the two schemes where the relay noise covariance matrix is $\bf K$ and $\bf K \odot I$ respectively are
\[\mbox{SNR}_{10}=\frac{{\bf \left| g^{T}Df\right|^2}P}{{\bf  g^{T}DKD^{*}g^{*}}+1}\]
\[\mbox{SNR}_{00}=\frac{{\bf \left| g^{T}Df\right|^2}P}{{\bf  g^{T}D(K\odot I)D^{*}g^{*}}+1}\]
Note that the power expended by the relays is the same in both the schemes.
When the relays have only local channel knowledge, the phase of the relays should be such that the phase of the forward and backward channels are
canceled. We therefore have
\[\mbox{SNR}_{10}=\frac{c}{b+a}~ \mbox{and}~\mbox{SNR}_{00}=\frac{c}{b}\]
where \[a=\sum_{i=1}^{N}\sum_{j=1~j\neq i}^{N}{\bf g}(i){\bf g^{*}}(j){\bf
 d}(i){\bf d}^{*}(j){\bf K}_{ij}\]
\[b=\sum_{i=1}^{N}|{\bf g}(i)|^{2}|{\bf  d}(i)|^{2}{\bf K}_{ii}\]
\[c=\left(\sum_{i=1}^{N}|{\bf d}(i){\bf g}(i){\bf f}(i)|\right)^2.\]
The term $a$ depends on both magnitude and phase of the channels while $b$ and $c$ depend only on the magnitude of the channels.
It can be shown that $\log(1+\mbox{SNR}_{10})$ is convex in $a$. Given the magnitude of the channels, the average of $a$ over channel phases is zero.
Therefore by Jensen's inequality
\[\mathbb{E}_{a,b,c}\left[\log (1+ \mbox{SNR}_{10})\right] \geq \mathbb{E}_{b,c}\left[\log \left(1+ \frac{c}{b}\right)\right]=\mathbb{E}_{b,c}\left[\log (1+ \mbox{SNR}_{00})\right] .\]

Now consider the case where the relays have no CSI. Here the relay
gain $\bf D$ is independent of $\bf f $ and $\bf g$. Taking
expectation of $\mbox{SNR}_{10}$ and $\mbox{SNR}_{00}$ over $\bf f$,
we have \begin{equation}\mathbb{E}_{{\bf
f}}[\mbox{SNR}_{10}]=\frac{{\bf
\parallel g^{T}D \parallel ^2}P}{{\bf
g^{T}DKD^{*}g^{*}}+1} \label{eqn:snr10_nocsi}\end{equation}
\begin{equation}\mathbb{E}_{{\bf f}}[\mbox{SNR}_{00}]=\frac{{\bf
\parallel g^{T}D \parallel ^2}P}{{\bf  g^{T}D(K \odot
I)D^{*}g^{*}}+1}\label{eqn:snr00_nocsi}\end{equation} where
$\mathbb{E}[{\bf ff^{\dagger}}]={\bf I}$. Observe that
(\ref{eqn:snr10_nocsi}) is dependent on the phase of $\bf g$ while
(\ref{eqn:snr00_nocsi}) is independent of phase of $\bf g$. From
Jensen's inequality it can be easily seen that
$\mathbb{E}[\mathbb{E}_{\bf f}[\mbox{SNR}_{10}]] \geq
\mathbb{E}[\mathbb{E}_{\bf f}[\mbox{SNR}_{00}]] $. Hence proved.
\hfill \QED

\subsection{Proof of Properties in Section \ref{SUBSECTION:PROPERTIES_THREE_HOP}} From
(\ref{input_output_three_hops}), the received signal at the
destination for the three-hop case is given by
\begin{eqnarray}
y={\bf g^{T}D_2HD_1f}x+{\bf g^{T}D_2HD_1n_1}+{\bf g^{T}D_2n_2}+n.
\end{eqnarray}
Let  ${\bf D_1}=\sqrt{P_1} {\bf D_1^{u}}$ and  ${\bf D_2}=\sqrt{P_2}
{\bf D_2^{u}}$ where ${\bf D_{i}^{u}}$ expends unit power at the
$i^{th}$ stage relays, $i \in \{1, 2\}$. Similarly let
$x=\sqrt{P_0}x'$ where $\mathbb{E}[|x'|^2]=1$. Now the effective
received signal at the destination is
\begin{eqnarray}
y'={\bf g^{T}D_2^{u}HD_1^{u}f}x'+\frac{{\bf
g^{T}D_2^{u}HD_1^{u}n_1}}{\sqrt{P_0}}+\frac{\bf
g^{T}D_2^{u}n_2}{\sqrt{P_1}}+\frac{n}{\sqrt{P_2}}.
\label{normalized_input_output_equation}
\end{eqnarray}
The modified power constraints are
\begin{equation}
\mbox{Tr}\left({\bf{D_1^{u}D_1^{u \dagger}}}\left({\bf
ff^{\dagger}}+\frac{{\bf I}}{P_0}\right)\right)=1,
\label{EQUATION_CONSTRAINT_FIRST_HOP_MOD}
\end{equation}
\begin{equation}
\mbox{Tr}\left({\bf D_2^{u}D_2^{u\dagger}}\left({\bf
HD_{1}^{u}ff^{\dagger}D_1^{u}H_2^{\dagger}}+\frac{\bf
HD_1D_1^{\dagger}H^{\dagger}}{P_0}+\frac{\bf
I}{P_1}\right)\right)=1. \label{EQUATION_CONSTRAINT_SECOND_HOP_MOD}
\end{equation}
\normalsize
\begin{enumerate}
\item{$P_2 \rightarrow \infty$}:\\
At very high $P_2$, the normalized input-output relation becomes
\begin{eqnarray}
y'&=&{\bf g^{T}D_2^{u}HD_1f}x'+\frac{{\bf
g^{T}D_2^{u}HD_1^{u}n_1}}{\sqrt{P_0}}+\frac{\bf
g^{T}D_2^{u}n_2}{\sqrt{P_1}}\\
&=&{\bf g^{T}D_2^{u}}\left({\bf HD_1^{u}f}x'+\frac{{\bf
HD_1^{u}n_1}}{\sqrt{P_0}}+\frac{\bf n_2}{\sqrt{P_1}}\right).
\end{eqnarray}
This represents a two-hop hop network with multiple antennas at the destination where ${\bf g^{T}D_2^{u}}$ is used as its receive combining vector. It is worth noting that maximal ratio combining (MRC) at the destination of the reduced network is capacity optimal. Let $\bf w_{{\small MRC}}$ be the optimal receive combining vector. Now, we need to show that there exists a diagonal matrix $\bf D_{2}^{u}$ in the original three-hop network such that ${\bf g^{T}D_2^{u}}= c {\bf w_{\small MRC}}$ for some constant $c$. This can be achieved when ${\bf D_2^{u}}(ii)=\kappa \frac{{\bf w_{\small MRC}}(i)}{{\bf g}(i)}$, where $\kappa$ satisfies the power constraint of the second stage relays given in (\ref{EQUATION_CONSTRAINT_SECOND_HOP_MOD}). (Here ${\bf D_2^{u}}(ii)$ denotes the $i^{th}$ element of the diagonal matrix ${\bf D_2^{u}}$.) Therefore we have
\begin{equation}\mathfrak{R}_{3}(P_0,{\bf
f},{\bf d_1}, P_1, {\bf H},{\bf d_2=d_2^{\circ}}, P_2=\infty, {\bf
g})=\mathfrak{R}_{2}( P_0,{\bf f},{\bf d_1}, P_1, {\bf H}
).\end{equation}

\item{$P_0 \rightarrow \infty$}:\\We prove this by using the previous
property and the reciprocity property of AF relay networks.
\begin{eqnarray}\mathfrak{R}_{3}(P_0=\infty,{\bf f},{\bf
d_1=d_1^{\circ}}, P_1, {\bf H},{\bf d_2}, P_2, {\bf
g})&=&\mathfrak{R}_{3}(P_2, {\bf g},\kappa_2{\bf d_2},  P_1, {\bf
H^{T}},\kappa_1{\bf d_1}=\kappa_1{\bf d_1^{\circ}}, P_0=\infty,{\bf
f})\label{eqn:recip1}\\&=&\mathfrak{R}_{2}(P_2, {\bf g},\kappa_2{\bf d_2},  P_1, {\bf
H^{T}})\label{eqn:prev_prop}\\&=&\mathfrak{R}_{2}(P_1, {\bf H},{\bf d_2}, P_2, {\bf
g})\label{recip2}
\end{eqnarray}
where (\ref{eqn:recip1}) follows from the reciprocity of multi-hop AF networks. (\ref{eqn:prev_prop}) follows from the preceding discussion on $P_2 \rightarrow \infty$. (\ref{recip2}) uses the reciprocity property again.

\item{$P_0, P_2 \rightarrow \infty$}:\\ For this scenario, we can
neglect $\bf n_1$ and $n$ from
(\ref{normalized_input_output_equation}) which results in
\begin{eqnarray}
y'&=&{\bf g^{T}D_2^{u}HD_1f}x'+\frac{\bf
g^{T}D_2^{u}n_2}{\sqrt{P_1}}\\
&=&{\bf g^{T}D_2^{u}}\left({\bf HD_1^{u}f}x'+\frac{\bf
n_2}{\sqrt{P_1}}\right) \label{eqn:ptp mimo}
\end{eqnarray}
with the power constraints being
\begin{equation}
\parallel{\bf D_1^{u}f}\parallel^2=1
\label{Eqn:ptp mimo_constraint1}
\end{equation}
\begin{equation}
\mbox{Tr}\left({\bf D_2^{u}D_2^{u\dagger}}\left({\bf
HD_{1}^{u}ff^{\dagger}D_1^{u}H_2^{\dagger}}+\frac{\bf
I}{P_1}\right)\right)=1. 
\end{equation}
(\ref{eqn:ptp mimo}) represents a point to point MIMO channel with $\bf D_1^{u}f$
and ${\bf g^{T}D_2^{u}}$ as its transmit precoding and receive
combining vectors respectively. Notice from (\ref{Eqn:ptp mimo_constraint1}) that the precoding vector
$\bf D_1^{u}f$ has unit norm. Therefore the transmit power for the channel in (\ref{eqn:ptp mimo}) is $P_1$.
Now let us consider the point to point MIMO channel $\bf Y=HX+N$. For any precoding vector $\bf u$ and receive combining vector $\bf v$ in this channel, there is an equivalent relay gain for the
first stage and second stage relays in the original network such that the two systems
are equivalent. That is, we can find $\bf D_1^{u}$ and $\bf D_2^{u}$ such that the following conditions are satisfied for some constant $c$: \[{\bf D_1^{u}f=u} ~\mbox{and}~{\bf  g^{T}D_2^{u}} =c {\bf v}\]
Since the precoding vector {\bf u} has to be unit norm, the power constraint for the first stage relays is automatically taken care of, while the constant $c$ takes care of the power constraint for the second stage relays. Therefore the original three-hop relay network can be reduced to a single hop point to point MIMO channel as given in the following.
\[\mathfrak{R}_{3}( P_0=\infty,{\bf f},{\bf d_1=d_1^{\circ}},
P_1, {\bf H},{\bf d_2=d_2^{\circ}}, P_2=\infty, {\bf
g})=\mathfrak{R}_{1}(P_1, {\bf H})\]

\item{$P_1, P_2 \rightarrow \infty$}:\\
At very high $P_1$ and $P_2$, the normalized input-output relation for the three-hop network
becomes
\begin{eqnarray}
y'&=&{\bf g^{T}D_2^{u}HD_1f}x'+\frac{{\bf
g^{T}D_2^{u}HD_1^{u}n_1}}{\sqrt{P_0}}\\
&=&{\bf g^{T}D_2^{u}HD_1^{u}}\left({\bf f}x'+\frac{{\bf
n_1}}{\sqrt{P_0}}\right).\label{eqn:simo}
\end{eqnarray}
The power constraints are
\begin{equation}
\mbox{Tr}\left({\bf{D_1^{u}D_1^{u \dagger}}}\left({\bf
ff^{\dagger}}+\frac{{\bf I}}{P_0}\right)\right)=1,
\label{EQUATION_CONSTRAINT_FIRST_HOP_MOD_SIMO}
\end{equation}
\begin{equation}
\mbox{Tr}\left({\bf D_2^{u}D_2^{u\dagger}}\left({\bf
HD_{1}^{u}ff^{\dagger}D_1^{u}H_2^{\dagger}}+\frac{\bf
HD_1D_1^{\dagger}H^{\dagger}}{P_0}\right)\right)=1. \label{EQUATION_CONSTRAINT_SECOND_HOP_MOD_SIMO}
\end{equation}

(\ref{eqn:simo}) represents a point to point SIMO system ${\bf
y=f}x+{\bf n}$ with transmit power $P_0$ and ${\bf
g^{T}D_2^{u}HD_1^{u}}$ as its receive combining vector. We now have
to prove that there exists diagonal matrices $\bf D_1^{u}$ and $\bf
D_2^{u}$ satisfying the power constraints such that ${\bf
g^{T}D_2^{u}HD_1^{u}}=c{\bf v}$ for some non-zero constant $c$. Here
$\bf v$ is the optimal MRC vector for the point to point SIMO
system. This can be achieved through the following: First, fix any
$\bf D_2^{u}$. Then, find ${\bf D_1^{u}}$ such that ${\bf
g^{T}D_2^{u}HD_1^{u}}=c{\bf v}$ where $c$ ensures compliance of
(\ref{EQUATION_CONSTRAINT_FIRST_HOP_MOD_SIMO}). Now scale $\bf
D_2^{u}$ to meet the constraint in
(\ref{EQUATION_CONSTRAINT_SECOND_HOP_MOD_SIMO}). Hence the original
three-hop network is equivalent to a point to point SIMO channel and
is represented as
\[\mathfrak{R}_{3}(P_0,{\bf f},{\bf d_1}, P_1, {\bf H},{\bf d_2=d_2^{\circ}},
P_2=\infty, {\bf g})=\mathfrak{R}_{1}( P_0,{\bf f}).\]

\item{$P_0, P_1 \rightarrow \infty$}:\\
By reciprocity, the proof for the previous case holds for this case
as well. \hfill \QED
\end{enumerate}

\bibliographystyle{ieeetr}
\bibliography{biblio}
\end{document}